# FUSE Observations of Interstellar and Intergalactic Absorption Toward the X-Ray Bright BL Lac Object Mrk 421

Blair D. Savage[1],  Bart P. Wakker[1],  Andrew J. Fox[1], and Kenneth R. Sembach[2]

## ABSTRACT

High-quality Far-Ultraviolet Spectroscopic Explorer (FUSE) observations at 20 km s$^{-1}$ resolution  of interstellar and intergalactic absorption from  910 to 1187 Å are presented for the X-ray bright BL Lac object Mrk 421.  These observations are supplemented with FUSE data for the distant halo stars BD +38  2182 and HD 93521 near the Mrk 421 line of sight, in order to obtain information about the distance to absorbing structures in the Milky Way toward Mrk 421.  The FUSE ISM observations provide measures of absorption by O VI and many other species commonly found in warm neutral and warm ionized gas including H I, C II, C III, O I, N I, N II, Fe II, and Fe III.  In this study  we consider the  O VI absorption between –140 to 165  km s$^{-1}$  and its relationship to the lower ionization absorption and to the strong absorption produced by O VII and O VIII at X-ray wavelengths. The O VI absorption extending from –140 to 60 km s$^{-1}$ is associated with strong low ionization gas absorption and originates in the Galactic thick disk / halo. This O VI appears to be produced by a combination of processes, including conductive interfaces between warm and hot gas and possibly cooling Galactic Fountain gas  and hot halo gas bubbles.   The O VI absorption extending from 60 to 165 km s$^{-1}$ has unusual ionization properties in that there is very little associated low ionization absorption, with the exception of C III.  This absorption is not observed toward the two halo stars, implying that it occurs in gas more distant than 3.5 kpc from the Galactic disk.   Over the 60 to 165 km s$^{-1}$ velocity range, O VI and C III  absorption have the same kinematic behavior.   N(O VI)/N(C III) = 10±3 over the 60 to 120 km s$^{-1}$  velocity range.  Given the association of O VI with C III, it is unlikely that the  high velocity O VI co-exists with the hotter gas responsible for the O VII and O VIII absorption.  The O VI positive velocity absorption wing might be tracing cooler gas entrained in a hot Galactic Fountain outflow.  The O VII and O VIII absorption observed by Chandra and XMM-Newton may trace the hot gas in a highly extended (~100 kpc) Galactic corona or hot gas in the Local Group.  The low resolution of the current X-ray observations (~750-900 km s$^{-1}$) and the kinematical complexity of the O VI absorption along typical lines of sight through the Milky Way halo make it difficult to clearly associate the O VI absorption with that produced by O VII and O VIII. A search for metal lines associated with the Ly $\alpha$  absorber at z = 0.01, which is situated in a Galactic void, was unsuccessful.

---

[1] Department of Astronomy,  University of Wisconsin, 475 N. Charter Street, Madison, WI 53706
[2] Space Telescope Science Institute, 3700 San Martin Drive, Baltimore, MD 21218

*Subject Headings:*  Galaxy:halo – ISM-atoms – ISM:clouds –  ultraviolet:ISM – intergalactic medium

## 1.  Introduction

The combination of ultraviolet  (UV) and X-ray absorption-line spectroscopy provides important diagnostic  information about  hot interstellar and intergalactic gas. At far-UV wavelengths  the lithium-like O VI doublet absorption lines at 1031.926 and 1037.617 Å are available, while at X-ray wavelengths  the helium-like and hydrogen-like transitions in O VII and O VIII are accessible at 21.602 and 18.970 Å, respectively.   In collisional ionization equilibrium these three ions peak in abundance at temperatures of $2.8 \times 10^5$ K (O VI), $7.9 \times 10^5$ K (O VII), and $2.2 \times 10^6$ K (O VIII) (Sutherland & Dopita 1993). Observations of O VI, O VII, and O VIII therefore provide unique information about hot plasma in the Milky Way disk and halo, in the Local Group, and in intervening intergalactic absorption-line systems.

The Far-Ultraviolet Spectroscopic Explorer (FUSE) produces spectra from 910 to 1187 Å with a spectral resolution of $\lambda/\Delta\lambda \sim 15,000$, corresponding to a velocity resolution (FWHM) of $\sim 20$ km s$^{-1}$ (Moos et al. 2000). With a small number of reflections and reasonably efficient detectors, FUSE provides an effective area of $\sim 50$ cm$^2$ at 1030!Å. (Sahnow et al. 2000). The observatories now available for X-ray spectroscopy of O VII and O VIII (Chandra and XMM-Newton) provide the modest velocity resolutions of 750 km s$^{-1}$ (Chandra) or 900 km s$^{-1}$ (XMM-Newton) and effective areas of $\sim 6$ cm$^2$ and 45 cm$^2$ near 20 Å, respectively (Paerels & Kahn 2003). Because AGNs and QSOs have much higher photon fluxes (per Hz) at 1000 Å compared to 20 Å, there now exists much more information about interstellar and intergalactic O VI with > 90 extragalactic objects providing good quality O VI measurements (Wakker et al. 2003) compared to only three extragalactic objects where zero redshift O VII and O VIII absorption lines have so far been reported in the literature at a significance level exceeding 4σ. The three include 3C 273, PKS 2155-304 and Mrk 421 (Nicastro et al. 2002; Fang, Sembach & Canizares 2003; Rasmussen, Kahn & Paerels 2003). While there have been studies linking O VI, O VII, and O VIII absorption for 3C 273 (Fang et al. 2003) and for PKS 2155-304 (Nicastro et al. 2002; Collins, Shull & Giroux 2004) there has been no comparable study of the UV and X-ray absorption toward the X-ray bright BL Lac object  Mrk 421.

Nicastro et al. (2002) believe the O VII and O VIII absorption and part of the O VI absorption are providing strong evidence for the existence of a warm-hot IGM filament with an extent of ~3 Mpc.    In contrast, Fang et al. (2003) note that while the WHIM filament explanation is exciting, it is not necessarily the correct explanation because an extended  (~100 kpc) hot corona of the Milky Way could also explain much of the observed  O VII and O VIII absorption.  Rasmussen et al. (2003) have placed approximate upper limits on the electron density ($n_e < 2x10^{-4}$ cm$^{-3}$) and scale length of the absorber (> 140 kpc) by interrelating X-ray absorption and emission observations. Although the density limit does not depend on the assumed oxygen abundance,  the scale length estimate depends inversely on  the oxygen abundance.   The length 140 kpc is for an assumed abundance of 0.3 solar.  These constraints are therefore consistent with either a very extended corona or halo containing O VII and O VIII or with a more extended intergalactic structure.  Determining the relative sizes of the galactic ISM and extragalactic IGM contributions to the strong  O VII and O VIII absorption near zero redshift will likely be a challenge for X-ray astronomy for many years.

In this paper we present high quality FUSE  observations of ISM and IGM absorption seen in the spectrum of the X-ray bright object Mrk 421 and two Galactic foreground stars over the wavelengths from 910 to 1187 Å.  These measurements are supplemented with a Goddard High Resolution Spectrograph (GHRS) spectrum  of Mrk 421 covering the wavelength region from 1221 to 1258 Å at a resolution of 15 km s$^{-1}$.  In our investigation we emphasize a study of  the possible relationships among  the absorption lines of O VI observed with FUSE and the lines of O VII and O VIII observed with XMM-Newton by Rasmussen et al. (2003).

In the following sections we first overview the Galactic and extragalactic structures found along the Mrk 421 line of sight (§2) and then discuss the FUSE and GHRS spectroscopic data handling (§3). In §4 and §5 we discuss the UV absorption at z = 0.0 and z = 0.01 found in the FUSE and GHRS observations. §6 and §7 contain discussions concerning the origin(s) of the O VI absorption near z = 0 and the relationships among the absorption produced by O VI, O VII, and O VIII. A summary of the investigation is given in §8.

## 2. The Line of Sight to Mrk 421

### 2.1. Galactic Structures

The UV and X-ray bright object BL Lac object Mrk 421 ($z_{em} = 0.03$) lies at high latitude in the Galactic anti-center direction $l = 179.83$ and $b = 65.03$. With a Galactic longitude close to 180, we would not expect Galactic rotation to influence the velocity of interstellar gas found along the line of sight to Mrk 421. Therefore, the measured velocity of absorption or emission by gas in this direction provides information about interstellar gas moving toward or away from the plane of the Milky Way.

Maps of 21 cm emission in this general region of the sky are shown in Figures 1a, b, c, and d and cover various ranges in H I velocity. Velocities given in this paper are all referenced to the Local Standard of Rest (LSR). For standard solar motion in the direction to Mrk 421, $v_{LSR} - v_{HELIO} = 2.7$ km s$^{-1}$. The maps in Figure 1 are based on observations from the Leiden-Dwingeloo H I survey (Hartmann & Burton 1997) with the grey scales and contours proportional to the measured H I column density in the velocity ranges –200 to –90 km s$^{-1}$ (Fig. 1a), –90 to –30 km s$^{-1}$ (Fig. 1b), 30 to 100 km s$^{-1}$ (Fig. 1c), and 100 to 200 km s$^{-1}$ (Fig. 1d). The antenna half power full width is 35′. Various H I high velocity clouds (HVCs) and intermediate velocity clouds (IVCs) are identified on the figures. The symbols identify Galactic directions for which O VI absorption line observations have been obtained for various extragalactic objects (including Mrk 421 from Wakker et al. 2003) and for the foreground stars HD 93521 (Widmann et al. 1998) and BD +38 2182 (Zsargo et al. 2003). The names, directions, and redshifts to these objects are listed in Table 1 along with information about the O VI absorption observed by FUSE and other satellites. The two foreground stars are at z-distances away from the Galactic plane, estimated to be 1.8 kpc (HD 93521; Howarth & Reid 1993) and 3.5 kpc (BD +38 2182; Ryans et al. 1997).

The neutral hydrogen 21 cm emission in the direction of Mrk 421 from the Green Bank 43 m radio telescope as a function of velocity is shown in the lower panel of Figure 2a. The H I observations have a resolution of ~1 km s$^{-1}$, an antenna beam size of 21′, and have been corrected for side-lobe contamination (see Murphy et al. 1996). Four components of H I emission are evident with component velocities, Gaussian line widths (FWHM), and H I column densities listed at the bottom of the Figure. Most of the H I emission occurs in components near –8 km s$^{-1}$ with N(H I) = (6.39±0.13)x10$^{19}$ cm$^{-2}$ and at –60 km s$^{-1}$ with N(H I) = (6.67±0.08)x10$^{19}$ cm$^{-2}$. The component at –60 km s$^{-1}$ is associated with the IVC IV26 (see Kuntz & Danly 1996; Wakker 2001). The total hydrogen column density in the direction to Mrk 421 obtained from the Green Bank observations is $1.5 \times 10^{20}$ cm$^{-2}$.  With a normal gas-to-dust ratio, this amount of H I implies a small amount of interstellar reddening.   We obtain  E(B-V) = 0.026 if we adopt the average value of N(H) / E(B-V) from Bohlin, Savage, & Drake (1979).  Even with the low value of N(H I) and E(B-V),  the far-UV spectrum of Mrk 421 is contaminated with numerous weak lines of  ISM H$_2$ absorption (see §4).

The line of sight to Mrk 421 passes through the Local  Bubble,   low velocity H I clouds near –10 km s$^{-1}$,  the intermediate velocity  cloud IV26 near –60km s$^{-1}$, the outer boundaries of HVC Complex M,  and the outer halo of the Galaxy.  Along this path many gas phases are sampled including the warm neutral medium (WNM),  the warm ionized medium (WIM), and the hot ionized medium (HIM), with scale  heights estimated to be ~0.5 kpc,  ~1 kpc, and ~2 to 4  kpc, respectively.  For a review of scale heights of the different phases of  the ISM see Savage (1995).   In addition,  a highly extended  ( > 100 kpc) corona  of hot gas and a more extended Local Group medium containing hot gas traced by O VII and O VIII  and possibly O VI may also be present.   Information about these numerous ISM structures and phases is given in Table 2 including information about the velocities and approximate distances to the structures.

Na I D-line absorption observations of several hundred local sight lines have been used to map the extent of the Local Bubble of low density gas (Sfeir et al. 1999).  At high galactic latitudes the Local Bubble appears to be partially open-ended  in both hemispheres with no well defined dense boundary for |z| < 0.2 kpc.  The distribution of detectable  extragalactic  EUV sources and the overall enhancement  of the soft 0.25 keV X-ray background emission at high latitudes supports the idea that the Local Bubble may open to chimney-like structures  into the halo in the north and south Galactic polar regions (Welsh et al. 1999). Although the boundaries of these chimney structures are not well defined, the direction to Mrk 421 does seem to be included in the direction of the chimney in the north (see Fig. 6 in Welsh et al. 1999).

The IVCs in directions near Mrk 421 have been investigated by Kuntz & Danly (1996) and Wakker (2001). For 7 IVCs that are part of the IV Arch, Ryans et al. (1997) find $0.4 \leq z \leq 3.5$ kpc while the summary of information given by Wakker (2001) constrains IV26 to lie in the range $z = 0.4$ to 1.8 kpc.

The IVCs in this general region of the sky have solar-like abundances for those elements normally unaffected by dust depletion effects (Wakker 2001). These IVCs are situated at substantial distances from the Galactic plane. With their abundances, distances, and infall velocities, these IVCs probably represent the cooled return flow of gas participating in a Galactic Fountain flow. The closest IVC to the Mrk 421 sight line is IV26, which is clearly seen in H I emission at $-61$ km s$^{-1}$ in the spectrum shown at the bottom of Fig. 2. The sight line to Mrk 421 may be tracing the cooler gas associated with the return flow of a Galactic Fountain at these velocities.

The direction of Mrk 421 also passes several degrees away from the outer boundaries of HVC Complex M, which consists of several clouds with velocities ranging from $-80$ to $-130$ km s$^{-1}$ (see Fig. 5 in Wakker 2001). Although there is no detected high velocity H I emission from Complex M in the direction of Mrk 421, UV absorption lines are extremely sensitive probes of low column density gas. At least some parts of Complex M are situated in the low halo with $z < 3.5$ kpc (Danly, Albert & Kuntz 1993) based on comparing the spectra of HD 93521 and BD +38 2182. At the closest positions to the Mrk 421 direction, Complexes MI, MII, and MIII have velocities of $-120$ , $-90$ , and

–90 km s$^{-1}$, respectively. The metallicity of Complex M (specifically cloud MI) has been estimated to be between 0.4 and 1.8 times solar (Wakker 2001), which is consistent with the idea that Complex M is the higher velocity part of the IV Arch and is tracing the cooled return flow of Galactic Fountain gas.

The FUSE survey of O VI in and near the Milky Way (Wakker et al. 2003; Sembach et al. 2003; Savage et al. 2003) revealed high positive velocity O VI absorption wings, sometimes extending to $v_{LSR} = 250$ km s$^{-1}$, for 21 of 100 extragalactic lines of sight. The distribution on the sky of objects showing these wings can be seen in Figure 17 of Sembach et al. (2003). The wings are most prevalent in the northern Galactic hemisphere with $l = 180$ to 360 and $b = 30$ to 80 which includes the direction to Mrk 421. The O VI profile for Mrk 421 measured by Wakker et al. (2003) has a weak wing extending from 100 to 185 km s$^{-1}$. A number of extragalactic objects near to Mrk 421 in the sky also have O VI positive velocity wings (see Table 1 and Fig. 1). Among the 13 extragalactic objects shown in Figure 1, five have the high positive velocity O VI wing and eight show no wings. For the two foreground stars near the direction to Mrk 421, O VI measurements have been reported by Jenkins (1978), Widmann et al. (1998), and Zsargo et al. (2003). The wing is not seen toward HD 93521 at z = 1.8 kpc in the Copernicus satellite observations (Jenkins 1978) or in the ORFEUS measurements (Widmann et al. 1998). BD +38 2182, at z = 3.5 kpc, observed with FUSE by Zsargo et al. (2003), also shows no evidence for the O VI wing. A more careful inspection of FUSE observations of these two stars in §4.3 also reveals the absence of the high positive velocity O VI wing. These measurements suggest that the wing in this general direction of the sky (see Figure 1) is produced in gas with z > 3.5 kpc. The wing may arise from outflowing Galactic Fountain gas, tidal debris associated with the Magellanic Stream, or Local Group gas (see Sembach et al. 2003). The new FUSE measurements for Mrk 421 allow a better study of the structure of the absorbing wing and the possible existence of other absorbing species at the velocity of the wing.

## 2.2. Extragalactic Structures

The path to Mrk 421 in regions beyond the Milky Way halo extends though possible Local Group gas on out to the $\sim$120 Mpc distance to the AGN ($z_{em} = 0.03$). The Local Group galaxies closest to the Mrk 421 line of sight are the E0 galaxy Leo II (UGC 6253, $l$ = 220.16 , $b$ = 67.23 , $v_{LSR}$ = −84 km s$^{-1}$), the dSph galaxy Leo I (UGC 5470, $l$ = 225.98 , b = 49.11 , $v_{LSR}$ = 288 km s$^{-1}$), and the irregular galaxy Leo A (UGC 5364, $l$ = 196.90 , $b$ = 52.42 , $v_{LSR}$ = 27 km s$^{-1}$). The velocities listed are from the National Extragalactic Database (NED). Note that for UGC 6253 (= Leo B and Leo II) inconsistent velocities are reported in the literature; $v_{LSR}$ = -84±5 km s$^{-1}$ by Falco et al. (1999) and 93±60 km s$^{-1}$ by de Vaucouleurs et al (1991). The impact parameters , $\rho$, between the Mrk 421 line of sight and the centers of these galaxies are $\rho$ = 59 kpc, 130 kpc, and 186 kpc, respectively.

Kravtsov, Llypin, & Hoffman (2002) have discussed possible observational signatures of intergalactic gas in the Local Supercluster region based on constrained hydrodynamical simulations of the flow of gas in the gravitational potential of the known structures within 100 Mpc of the Milky Way. Assuming an oxygen abundance of 0.3 solar in the IGM, the simulations predict the existence of a Local Supercluster gas filament containing O VI with N(O VI) > $10^{14}$ cm$^{-2}$ covering $\sim$45% of the Galactic sky with b > 30 . Although the expected kinematical properties of this gas are uncertain, it is

doubtful that the absorption produced by Local Supercluster gas situated well beyond the

Local Group would coincide with the O VI observed by FUSE at the modest velocities of

$|v_{LSR}| < 165$ km s$^{-1}$.

Beyond the Local Group, the path to Mrk 421 passes near the galaxies and galaxy

groups whose positions are shown in Figure 3.  This plot shows all galaxies in this region

that are listed in the NASA/IPAC Extragalactic Database.  The symbol shapes  are coded

according to redshift, while the symbol areas are proportional to the physical size of each

galaxy.   Galaxy groups in  the catalog of Geller & Huchra (1983) have boundaries  that

lie ~ 3  to 8  from the line of sight to Mrk 421, including GH75 (748 km s$^{-1}$), GH67

(1697 km s$^{-1}$), GH84 (1882 km s$^{-1}$), GH86 (2680 km s$^{-1}$), and GH73 (6851 km s$^{-1}$).

Table 3 lists the known galaxies beyond the Local Group closest  to the Mrk 421 line of sight.  We give galaxy identification,  $l$,  $b$,  galaxy type when known, V magnitude,

radial velocity, distance, galaxy diameter,  and  the impact parameter,  $\rho$.  The size and

distance estimates assume $H_o = 71$ km s$^{-1}$ Mpc$^{-1}$.  The 8 galaxies with v ranging from 603

to 756 km s$^{-1}$ and $\rho$ from 0.52 to 1.38 Mpc are part of the very nearby Leo Spur grouping of galaxies with v =  620 $\pm$160 km s$^{-1}$ identified as grouping 15 by Tully (1988).  This close-by galaxy grouping  has members covering the entire field shown in Figure 3.

McLin et al. (2002) have searched  for faint galaxies associated with low redshift Ly

$\alpha$ absorbers in voids, which they define as regions having no bright galaxy ($M_B \leq$ -17.5)

closer than 2 Mpc.  They included in their study the Ly $\alpha$  absorber toward Mrk 421 at

cz= 3035 km s$^{-1}$, with a restframe equivalent  width $W_r = 86\pm15$ mÅ  (Penton, Shull &

Stocke 2000).   McLin et al. (2002) find no galaxies within $\pm$300 km s$^{-1}$  of this absorber down to an apparent blue magnitude limit of 18.7, which corresponds to an absolute blue magnitude limit of –14.5. The radius of the region covered in the search at the distance of the absorber is 252 kpc and the completeness limit is estimated to be 95%.

## 3. **Observations**

### 3.1. FUSE Observations of Mrk 421

The basic properties of FUSE and its in-orbit performance are discussed by Moos et al. (2000) and Sahnow et al. (2000). The FUSE observations of Mrk 421 analyzed in this investigation are listed in Table 4, where we give the observation date, the FUSE program ID, the observation ID, the FUSE aperture, and the actual integration time. The 21.8 ksec integration obtained in 2000 was part of the FUSE survey of O VI in the Milky Way halo and beyond (Wakker et al. 2003). That observation produced a spectrum with S/N = 16.5 per 20 km s$^{-1}$ resolution element near 1030 Å. With the additional 62 ksec of integration obtained in early 2003 on Mrk 421, it is possible to substantially improve on the S/N of the measurements published by Wakker et al. (2003), and to study the relationships among the many absorbers appearing in the FUSE spectra that extend from 910 to 1187Å. All the observations of Mrk 421 listed in Table 4 were obtained in the time- tagged mode with Mrk 421 centered in the large (30″ x 30″) aperture of the LiF1 channel.

The spectrum extraction and calibration techniques we employed are similar to those discussed by Wakker et al. (2003) except that all the measurements were processed with the newer version 2.4.0 of the data calibration pipeline, which provides a more reliable wavelength calibration. The multiple spectra obtained for each of the three segments (LiF1A, LiF2A, and SiC2A) were combined by fitting Gaussians to the profiles of the prominent ISM features in each of the observations, separately. For each segment an average velocity was determined and compared with the expected velocities of −10 and −61 km s$^{-1}$ based on the H I 21 cm emission profile. A final single spectrum was produced by combining the measurements from the different spectrograph/detector segments in the overlapping wavelength regions. The S/N ratio per 20 km s$^{-1}$ resolution element in the final combined spectra have values of ~12 to 19 from 930 to 1000!Å and ~25 to 20 from 1030 to 1150 Å. For a detailed discussion of our approach for determining errors in equivalent widths and other quantities see Wakker et al. (2003). The errors on the equivalent widths for Mrk 421 are obtained by a quadrature addition of the continuum placement error, the statistical error, and an estimated 5 mÅ fixed pattern noise error.

## 3.2 GHRS Observations of Mrk 421

The GHRS spectrum of Mrk 421 is important for identifying low redshift intervening H I Ly α absorption lines. It also provides measures of the ISM lines of S II λλ1250.584, 1253.811 and a limit on ISM N V λλ1238.821, 1242.804. The GHRS spectrum was obtained as part of a low redshift IGM program lead by J. Stocke. The spectrum was obtained on 1 February 1995 with the G160M grating using the large 1.74″x1.74″ entrance aperture and the G160M grating. The spectrum covers the wavelength region from 1221 to 1258 Å and has a resolution of 19.7 km s$^{-1}$ (FWHM), which is comparable to the FUSE resolution. For details of the post-COSTAR performance properties of the GHRS see Robinson et al. (1998). The GHRS spectrum was extracted and velocity calibrated by using the standard calibration pipeline provided through the Multimission Archive a the Space Telescope Science Institute (MAST). The S/N per 20 km s$^{-1}$ resolution element obtained in the final extracted spectrum near 1250 Å is 22.

### 3.3 FUSE Observations of HD 93521 and BD +38 2182

The two halo stars HD 93521 and BD +38 2182 lie 3.2 and 3.3 degrees away from the line of sight to Mrk 421 at distances of 1.8 and 3.5 kpc away from the Galactic plane, respectively (see Table 1). To study the stratification of O VI away from the Galactic plane in this general direction it is of interest to carefully compare the O VI absorption toward these two stars with the absorption toward Mrk 421. We have therefore obtained the FUSE observations of these two stars listed in Table 4 and have extracted the spectra to compare the stellar and extragalactic measurements of the O VI absorption. For these two stars spectra were obtained in the histogram mode with the large (LWRS; 30"x30") science aperture for BD +38 2182 and the medium (MDRS; 4"x20") and small (HIRS; 1.25"x20") apertures for HD 93521 (see Table 4).

For BD+38 2182 our spectral extraction procedures are identical to those used for Mrk 421. In the region of O VI, the final spectrum we used was the one obtained in the LiF1A detector segment, because this segment produces spectra with higher S/N and resolution than the other segments (see Wakker et al. 2003).

HD 93521 is so bright in the far-UV it was not possible to safely observe it with the LiF1A and LiF2B detector segments. The observations were therefore obtained with the SiC1A and SiC2B segments. Since the spectral resolution is somewhat higher in the SiC1A segment we have chosen to extract information on the O VI absorption from those measurements, although both channels give very similar results. The HD 93521 observations were obtained using the FUSE focal plane-split observation procedure to reduce the effects of detector fixed-pattern noise. In the final processing of the observations the individual integrations were first aligned before co-addition. We thank J. Kruk for providing the extracted HD 93521 observations. The full details of the special steps taken will be given in a future paper on D/H in the IVCs toward HD 93521.

### 4. Far-UV Absorption at z = 0.0

The FUSE absorption line observations of Mrk 421 near zero redshift as a function of LSR velocity are shown in Figures 2a, 4, and 5. The FUSE observations of HD 93521 and BD +38 2182 are shown in Figures 2b and 2c. The various panels in Figures 2a, b, and c display from top to bottom the FUSE spectrum from 1015 to 1049 Å, absorption profiles of O VI $\lambda\lambda$1031.926, 1037.617, apparent column density, $N_a(v)$, profiles for the O VI $\lambda\lambda$1031.926, 1037.617 lines, absorption profiles for C II $\lambda$1036.337, Si II $\lambda$1020.699, Ar I $\lambda$1048.220, and radio observations of the H I emission brightness temperature measured by the Green Bank NRAO 43-m telescope for Mrk 421 and the Jodrell Bank 76 m telescope for the two stars. Various other species that contribute to the absorption seen in the ±450 km s$^{-1}$ wide panels are indicated below each spectral panel. Lines marked as R0, R1, R2, R3, R4, P1, P2, and P3 refer to transitions in H$_2$.

For all three objects the blending confusion produced by H$_2$ and other absorbers is serious for the region around the O VI $\lambda$1037.617 line, but hardly affects the O VI $\lambda$1031.926 line, which has H$_2$ 6-0 P(3) absorption from the 1031.191 Å line appearing at $v_{LSR} = -214$ km s$^{-1}$ and from the H$_2$ 6-0 R(4) 1032.356 Å line occurring at $v_{LSR}= 125$ km s$^{-1}$. The strength of these H$_2$ lines near the O VI $\lambda$1031.926 absorption can be modeled with reference to the numerous H$_2$ lines appearing at other wavelengths in the FUSE spectrum (see Wakker et al. 2003 for the details). Note that the format of Figure 2 is the same as for the spectra illustrated for 100 AGNs and two stars in the FUSE O VI survey of Wakker et al. (2003).

The solid lines placed through the top of the spectra in the various spectral panels of Figure 2 show the continuum we have adopted for the analysis of the observations, including the effects of H$_2$ absorption, when necessary . The continuum placement for objects like Mrk 421 is quite simple because the intrinsic emission spectrum is dominated by non-thermal emission which produces a relatively flat and featureless continuum. For the stellar observations of HD 93521 and BD +38 2182 the continuum placement is more uncertain. For detailed discussions of the uncertainties of the continuum placement problems for studying O VI absorption toward hot stars see Zsargo et al. (2003).

## 4.1. Mrk 421 Absorption Line Kinematics

The O VI $\lambda$1031.926 absorption line in the spectrum of Mrk 421 is strong and extends from –140 to 165 km s$^{-1}$. In contrast, the 21 cm H I emission in this direction shown at the bottom of Figure 2a is confined to a narrower velocity range from –100 to 50 km s$^{-1}$ and breaks up into distinct Gaussian components, with the velocities, brightness temperature amplitude, full line width at half maximum (FWHM), and H I column density, N(H I), listed at the bottom of Figure 2a. The highest column density components of H I at –60 km s$^{-1}$ and –8 km s$^{-1}$, with N(H I) = (66.7±0.8)x10$^{18}$ cm$^{-2}$ and (63.9±1.3)x10$^{18}$ cm$^{-2}$, are also seen in the FUSE absorption measurements of Ar I λ1048.220 and Si II λ1020.699.

It is possible there may be several more low velocity components in the line of sight, all

blended together.   The FUSE observations of these absorption lines are substantially

degraded by the 20 km s$^{-1}$   (FWHM) resolution of the FUSE spectrograph.   Absorption in

the very strong and saturated C II λ1036.337 line extends from –150 to 60  km s$^{-1}$.    The

O VI absorption from 60 to 165 km s$^{-1}$ occurs in a velocity region where there is no C II

λ1036.337  absorption.   Numerous other absorption lines appearing in the FUSE

spectrum are illustrated in Figures 4a, b, c.

The interstellar absorption lines definitely identified in the FUSE spectra include:

The H I  Lyman series to Ly μ λ917.181, C II λ1036.337, C II* λ1036.791, C III

λ977.020, many N I lines,  N II λ1083.990, O I λλ921.860, 924.952, 929.517, 936.630,

948.686, 950.885, 971.738, 976.448, 988.773, 1039.230,  O VI λλ1031.926, 1037.617,

Si II λλ1020.699,  P II λλ1152.818,  S III λ1012.502, Ar I λλ1048.220, 1066.660, Fe II

λλ940.192, 1055.262, 1063.176, 1081.875, 1096.877, 1112.048, 1121.975, 1125.448,

1133.665, 1142.366, 1143.226, 1144.938,  and Fe III λ1122.526.   In addition, there are

numerous weak  $H_2$ lines near $v_{LSR}$ = -10 km s$^{-1}$  arising from the J = 0, 1, 2, and 3

rotational states, respectively.   The strongest $H_2$ lines detected at > 2σ  significance have equivalent widths ranging from 15 to 60 mÅ.   Seven $H_2$  lines arise from the J = 0 level,

17  from the J = 1 level,  10 from the J= 2 level, and 8 from the J = 3 level.   No $H_2$

absorption is seen out of levels with J ≥ 4 or at the  - 61 km s$^{-1}$  velocity of the IVC

toward Mrk 421 (also see Richter et al. 2003).   The ISM lines definitely detected in the

GHRS spectrum extending from 1221 to 1258 Å  include S II  λλ1250.584, 1253.811.

ISM absorption by N V λλ 1238.821, 1242.804 is not detected.

The lines of Ar I, Fe II, Si II, Fe III, and S III displayed in Figures 2a and 4 show the two principal component structure of the 21 cm emission. There are many lines where the IVC at –60 km s$^{-1}$ is recognizable as a component separated from the low velocity absorption near –8 km s$^{-1}$. These include Si II $\lambda$1020.699, P II $\lambda$963.801, S II $\lambda\lambda$1250.584, 1253.811, Ar I $\lambda\lambda$1048.220, 1066.660, and most of the Fe II lines. For the stronger low ionization absorbers, including almost all of the lines of C II, O I, and N I, the absorption by the two components blend together.

In an independent investigation, Lehner, Wakker & Savage (2004) have studied the low and intermediate ionization species in FUSE spectra of extragalactic sources, in order to determine the distribution of Fe II, Fe III, P II, and the cooling of halo gas through the C II fine structure transition at 157.7 $\mu$m, which originates from the level producing the C II* $\lambda$1036.791 absorption line. Therefore, in this paper we concentrate on the O VI absorption and its kinematical relationships to the other phases of the absorbing gas as traced by the weak and strong lines produced by these other ion lines.

At high negative velocity, a component near –125 km s$^{-1}$ is detected in C II $\lambda$1036.337, N II $\lambda$1083.990, O I $\lambda$988.773, and possibly in S III $\lambda$1012.502. The absorber is also seen in the higher H I Lyman series lines down to Ly $\lambda$918.129 and in C III $\lambda$977.020, where blending with O I $\lambda$976.448 must be allowed for. Mrk 421 passes ~4 east of HVC MI which has $v_{LSR}$ = –120 km s$^{-1}$. The FUSE observations are likely detecting the low column density outer boundaries of HVC MI in H I, C II, C III, and S III. The observations only provide rough information about the column densities in the various ions detected in this HVC. The 21 cm observations in the direction of Mrk 421 do not reveal the HVC, implying N(H I) < $2x10^{18}$ cm$^{-2}$. The H I column density in this –125 km $s^{-1}$ absorber is estimated to be N(H I) ~ 4.8 (+6.9, -2.8) )x$10^{18}$ cm$^{-2}$ from a component fit to the H I Lyman series (see §4.2). The negative velocity wing of the O VI λ1031.926 absorption extends to the same negative velocity as C II, suggesting a kinematical association of the absorption by the low and highly ionized gas in this HVC.

In Figure 4a we show how the O VI λ1031.926 profile behavior relates to that for a very strong low ionization line (C II λ1036.337), for a very strong intermediate ionization line (C III λ977.020), for a moderately strong intermediate ionization line (N II λ1083.990), two weak intermediate ionization lines (S III λ1012.5.2 and Fe III λ1121.975), and for extremely strong to moderate strength H I Lyman series lines. The H I absorption measurements shown are from orbital night-only observations in order to reduce the effects of geocoronal H I emission, which is marked with the ⊕ symbol in the various panels. This atmospheric emission in the night only data is strongest for Ly β, weakens for the higher members of the Lyman series, and is basically absent in the lines below 926Å. In the full orbit data, the emission is present in all the H I lines down to 917 Å. The O VI absorption from –140 to 60 km s$^{-1}$ covers the velocity range spanned by species tracing weakly and moderately ionized gas. In contrast, the O VI absorption wing extending from 60 to 165 km s$^{-1}$ is not seen in the strong line of C II, but appears to be related to the weak absorption wing seen in C III from 60 to ~165 km s$^{-1}$. Absorption by H I is not seen at these high velocities, except for Ly β where the apparent absorption is from the damping wing of strong H I absorption at low velocity (see the HI profile fit in Fig. 4a).

*4.2. Column Densities*

The FUSE resolution of 20 km s$^{-1}$ is adequate to nearly fully resolve the ISM absorption produced by O VI because at T ~ 3x10$^5$ K, the temperature at which O VI peaks in abundance under conditions of collisional ionization equilibrium, the thermal Doppler broadening of O VI corresponds to a FWHM of 29 km s$^{-1}$. Since the O VI line is nearly fully resolved, the most straightforward way to determine column density of O VI over various velocity ranges is to use the apparent optical depth method (Savage & Sembach 1991).

The fourth set of measurements from the top in Figure 2a compares the apparent column density profiles for the strong line (heavy line in the figure) and weak line (light line in the figure) of O VI for Mrk 421. Near the line center, where the weak O VI λ1037.617 line profile is not affected by blending from various other species, we see that the strong and weak line values of $N_a(v)$ agree very well. This good agreement confirms that the O VI absorption is nearly fully resolved and that the $N_a(v)$ profiles are hardly affected by line saturation. We can therefore determine reliable values of N(O VI) integrated over different velocity ranges by integrating over the $N_a(v)$ profile for the 1031.926 Å line. We note that the significantly higher S/N of these new observations for Mrk 421 has removed concerns about the line saturation problems that accompanied the observations published by Wakker et al. (2003) based on the first 21.8 ks integration (see Table 4). In the earlier observation it was found that N(O VI λ1037.617)/N(O VI λ1031.926) integrated over the velocity range from –100 to 100 km s$^{-1}$ was 1.60±0.18, which suggested substantial unresolved line saturation, since the weaker absorption line was producing a larger value of the column density than the stronger absorption line.

However, by combining these old observations with the new measurements we instead obtain 1.08±0.07 for this ratio. The new observations are consistent with little or no line saturation for the Mrk 421 O VI absorption. The difference between the old and new measurements is probably due to a combination of an improvement in the reliability of the continuum placement and a reduction of the statistical noise in the observations.

The fourth panel of Figure 2c for HD 93521 compares the strong and weak line values of $N_a(v)$ for O VI for this stellar line of sight. Again there is good agreement which implies there is little or no unresolved line saturation in the O VI absorption toward HD 93521. However, in the case of BD +38 2128, the major amount of stellar and interstellar line blending near the weaker O VI $\lambda$1037.617 line makes the O VI doublet line intercomparison unreliable (see Fig. 2b). For this star we therefore rely entirely on the values of $N_a(v)$ derived from the strong $\lambda$1031.926 line of the O VI doublet after removing the effects of $H_2$ absorption. The FUSE observations of Savage et al. (2003), Zsargo et al. (2003), and Howk et al. (2003) imply that unresolved saturation usually does not affect the measures of $N_a(v)$ for O VI toward halo stars.

Equivalent widths and column densities for the various absorption lines of interest for the interpretation of the highly ionized oxygen absorption toward Mrk 421 are given in Table 5, where we list the ion, rest wavelength in Å, values of log(f$\lambda$) from Morton (2003), the velocity range for the equivalent width integration, equivalent width, $W_\lambda$ in mÅ, along with the errors, the values of logN including the errors. The column densities and limits were derived by the methods summarized in the notes to Table 5. Our methods for deriving the equivalent width and column density errors are described by Wakker et al. (2003). The error listed include the statistical error, the continuum placement error, and the fixed pattern noise error added in quadrature. We estimate the fixed pattern noise error to be 5 mÅ for the spectral regions covered by these observations.

The H I absorption in the vicinity of the Lyman lines was modeled to produce good fits to the night time H I measurements as shown in Fig. 4a. Even though these are night time-only observations, there is still strong geocoronal H I emission, particularly for Ly β λ1025722, Ly γ λ972.537, and Ly δ λ949.743. The H I absorption model fit includes three components with the values of v (km s$^{-1}$), $b$(km s$^{-1}$), and N(H I) (cm$^{-2}$) listed in Table 6. The 21 cm emission observations (Fig. 2a) and the FUSE O I absorption observations (Fig. 4) were use to guide the component fitting process by determining the likely velocities of the absorbing components. The final fit required components at –125, -60 and –8 km s$^{-1}$. At the FUSE resolution of 20 km s$^{-1}$ the strong component at –8 km s$^{-1}$ subsumes the weak components seen at –28 km s$^{-1}$ and 18 km s$^{-1}$ identified in the 21 cm emission profile. The approximate errors assigned to the b values and column densities in Table 6 are estimated by varying the fit parameters and determining the change in the value of the reduced chi square for the fit. The component at –125 km s$^{-1}$ is not seen in 21 cm emission but is detected in C II λ1036.337 absorption. No other components are required to model the H I absorption. In particular, there appears to be no additional H I absorption that might be associated with the highly ionized C III and O VI absorption in the velocity range from 60 to 165 km s$^{-1}$.

### 4.3. Absorption Distance Constraints Provided by HD 93521 and BD +38 2182

The two foreground stars, HD 93521 and BD +38 2182, provide interesting information about the possible distances away from the galactic plane of the O VI absorption seen toward Mrk 421. HD 93821 lies at z = 1.8 kpc while BD +38 2182 is at z = 3.5 (see Table 1 and §2.1). While both stars lie ~ 4 from the direction to Mrk 421, 4 out of 5 extragalactic objects in the Galactic anti-center directions bounded by $l = 170$ to 190 (see Fig. 1) display O VI absorption extending to large positive velocities while the two stars do not display the high positive velocity absorption.

Figure 5 displays $N_a(v)$ profiles for the O VI $\lambda1031.926$ line for the directions to Mrk 421, BD +38 2182, and HD 93521. Values of logN(O VI) and the logarithmic column density perpendicular to the Galactic plane log[N(O VI)sin|b|] integrated over different velocity ranges for these three objects are listed in Table 7.

The value of log[N(O VI)sin|b|] = 14.34±0.02 toward Mrk 421 for the velocity range from −140 to 60 km s$^{-1}$ is typical of sight lines through the thick O VI disk of the Galaxy in the north Galactic polar region. For 26 extragalactic lines of sight with $b =$ 60 to 90 , Savage et al. (2003) found log[N(O VI)sin|b|] to have median and average values of 14.38 and 14.37 with a standard deviation of 0.15 dex. The perpendicular column density N(O VI)sin|b| = $2.2 \times 10^{14}$ cm$^{-2}$ toward Mrk 421 implies an exponential scale height for the gas of 4.2 kpc, assuming a mid-plane O VI density of n(O VI) = $1.7 \times 10^{-8}$ cm$^{-3}$ from the FUSE O VI disk survey of Jenkins et al. (2001). With a scale height of 4.2 kpc, the values of log[N(O VI)sin|b|] we would expect to see toward HD 93521( z = 1.8 kpc) and BD +38 2182 (z = 3.5 kpc) are 13.92 and 14.13 respectively. These numbers are close to the values of 13.96 and 14.12 that are observed (see Table 7).

The O VI observations are consistent with the majority of the O VI absorption from –140 to 60 km s$^{-1}$ being associated with the thick disk of the Milky Way.  However, there is a considerable amount of irregularity associated with the thick disk O VI absorption (see Savage et al. 2003 and Howk et al. 2002).  Therefore, we cannot rule out the possibility that a substantial fraction (~50%) of the O VI absorption we attribute to the thick Galactic disk in fact occurs in a more distant region of the halo or the Local Group.

The O VI absorption toward Mrk 421 is much broader than that observed toward the halo stars (see Fig. 5).  The broad  negative and positive velocity O VI absorption toward Mrk 421  therefore probably occurs in gas with z > 3.5 kpc.  Gas in the positive velocity wing of absorption toward Mrk 421 from 60 to 165 km s$^{-1}$  with logN(O VI) = 13.66±0.06 is  of particular interest because of its unusual ionization characteristics, with only C III and O VI detected.  Toward the two halo stars BD+38$^{o}$2182 and HD 93521 the 3σ limits for O VI absorption from 60 to 165 km s$^{-1}$ are logN(O VI) $\leq$ 13.22  and 13.28, respectively (see Table 7).  These limits were obtained by making plausible assumptions about errors in the  stellar continuum placement over the velocity range from 60 to 165 km s$^{-1}$ in Figs. 2b and 2c.

## 5.  Far-UV Absorption at z = 0.01

The HST/GHRS observations of Penton et al. (2000) for Mrk 421 revealed  H I Ly α absorption  with a rest-frame equivalent  width of 86±15 mÅ at  cz = 3035±6 km s$^{-1}$ corresponding to  z = 0.01012±0.00002.  This absorber has an H I column density of logN = 13.25±0.04  and  $b$ = 29.5±5.6 km s$^{-1}$, which is inferred from the observed line profile, corrected for instrumental blurring (see Table 8).  It is of interest to look in the FUSE wavelength range for higher H I Lyman series lines and for metal lines that might be associated with this Ly $\alpha$ absorber. The search is particularly interesting since the Ly $\alpha$ absorber occurs in a void (see §2.2 and McLin et al. 2002).

Our reductions of the GHRS Ly $\alpha$ observation yields $W_\lambda$(rest) = 82±6±7 mÅ at cz = 3022 km s$^{-1}$ in the LSR reference frame. We found it necessary to shift the spectrum produced by MAST by +15 km s$^{-1}$ in order to have the velocities from the ISM S II lines agree with those from the H I 21 cm emission profile and the Ar I $\lambda$1048.220 absorption. Without that shift the value of cz we obtain for the Ly $\alpha$ is 3007 km s$^{-1}$. The following discussions utilize our values of the velocity and redshift of this absorber.

Figure 6 shows velocity profiles for important ions we searched for near cz = 3022 km s$^{-1}$ or z = 0.010073 in the FUSE observations for Mrk 421. The top panel shows the GHRS Ly $\alpha$ observation. In the next panels H I Ly $\beta$ is contaminated by Galactic C II $\lambda$1036.337 absorption, and H I Ly $\gamma$ is not detected. We therefore can not confirm the Ly $\alpha$ identification by finding a higher Lyman series absorber. However, it is very unlikely that such a strong line could be anything but Ly $\alpha$.

In our search for metal lines associated with this system we find that C II $\lambda$1036.337 is not detected. C III $\lambda$977.020 is contaminated by the ISM H$_2$ R(1) line at 986.764 Å. The fitted H$_2$ profile shown in Figure 6 is for an H$_2$ line at 986.796 Å corresponding to $v_{LSR}$ = -10 km s$^{-1}$ with $W_\lambda$ = 55.5 mÅ, which is based on several other J = 1 H$_2$ lines with similar values of f$\lambda$. The fit explains most of the absorption in the region of C III $\lambda$977.020 and we can only place an upper limit on the amount of residual absorption that could be C III. The 3$\sigma$ limit is $W_\lambda$ (rest) = 33 mÅ which corresponds to logN(C III) < 12.72, assuming no line saturation. Continuing with the panels in Figure 6 we find that N V $\lambda\lambda$1238.821 and 1242.804 are absent,  O VI $\lambda$1031.926 is absent,  and  O VI $\lambda$1037.617 is contaminated with ISM Ar I $\lambda$1048.220.  All of these results are summarized in Table 8 where restframe equivalent widths or 3$\sigma$ limits are listed for the various ions along with 3$\sigma$ limits to the metal line column densities.

The value of logN(H I) = 13.25±0.04 and the 3$\sigma$ limit on O VI,  logN(O VI) < 13.18, implies N(O VI)/ N(H I) <  0.85.  When O VI is observed in QSO absorption line systems the values of N(O VI)/N(H I) span a very large range from ~0.1 to 10 (Shull 2003).  The value of *b* for the Ly$\alpha$ absorber of 30 km s$^{-1}$ (corrected for the GHRS instrumental blurring) implies T < 5.4x10$^4$ K.  Therefore the H I does not exist in gas hot enough to collisionally ionize O VI.

The absence of O VI is interesting since Nicastro (2004) claims the O VII  He$\alpha$ $\lambda$21.602 line  is detected in this system at 21.852Å corresponding to z = 0.01157  or cz = 3471 km s$^{-1}$ with N(O VII) = (1.2±0.3)x10$^{15}$ cm$^{-2}$.  The O VII column density assumes no line saturation so the true value could be larger.  In collisional ionization equilibrium O VII peaks in abundance at logT(K) = 5.9.  At this temperature N(O VI)/N(O VII) = 5.6x10$^{-3}$, implying the amount of associated O VI would be  ~6.7x10$^{12}$ cm$^{-2}$.  If the line identification is correct,  the O VII is displaced to positive velocity by  ~450 km s$^{-1}$ from the Ly $\alpha$ absorber.  This velocity offset is uncertain because of the low spectral resolution of Chandra.  Note that the FUSE measurements shown for the O VI  $\lambda$1031.926 absorption line in the velocity range 2500 to 4000 km s$^{-1}$ in Figure 6 are from the LiF1A and  the LiF2B detector segments.  The LiF1A segment observations are contaminated by a  broad detector flaw extending over the velocity range from 3210 to 3425 km s$^{-1}$which corresponds to the wavelength range from 1043.0 to 1043.7 Å.  The feature we identify as a detector flaw can't possibly be O VI  $\lambda$1031.926 at 3300 km s$^{-1}$ because the corresponding weaker member of the O VI doublet at  $\lambda$1037.617 is not detected.  The detector flaw is due to a dead spot. It is obvious in the co-added flat field image for the LiF1A detector segment. The flaw can also be seen in published FUSE spectra. For example, see the feature at 1043 Å in Figure 2g of Sembach et al. (2004). In the O VI $\lambda 1037.617$ panel, the weak fitted feature near 3400 km s$^{-1}$ is the expected position and strength of the ISM H$_2$ $\lambda 1049.367$ R(0) line. That line is not detected while the stronger ISM H$_2$ $\lambda 1049.960$ R(1) line near 3600 km s$^{-1}$ is detected. No O VI $\lambda 1031.926$ or O VI $\lambda 1037.617$ absorption is evident at the expected velocity of ~3450 km s$^{-1}$. We conclude there is no obvious O VI absorption over the velocity range from 2500 to 4000 km s$^{-1}$ that can be associated with the claimed O VII absorption. The rest frame $3\sigma$ equivalent width upper limits for possible O VI $\lambda 1031.926$ absorption in 100 km s$^{-1}$ wide features is 27 mÅ for v = 2500 to 4000 km s$^{-1}$, but excluding the region from 3150 to 3500 km s$^{-1}$ affected by the LiF1A detector defect (see Fig. 6). Over the velocity range from 2500 to 4000 km s$^{-1}$, the 3 sigma rest frame equivalent limit derived from the LiF2B observations is 34 mÅ. These equivalent width limits correspond to $3\sigma$ column density limits of logN(O VI) < 13.33 and <13.43, respectively.

## 6. Properties of O VI Absorption near z = 0.0

### 6.1. O VI Thick disk absorption.

The Milky Way has an atmosphere of gas extending into the halo with an extension that is dependent on the gas phase. The gas in the hot phase traced by O VI shows the greatest extension. We refer to this atmosphere as the thick disk of the Milky Way. Thick disk O VI absorption toward Mrk 421 extending over the velocity range from –140 to 60 km s$^{-1}$ is typical of other lines of sight in the north Galactic polar direction. We estimate the exponential scale height of the gas containing O VI in the direction of Mrk 421 to be ~ 4.2 kpc (see § 4.3). The O VI thick disk absorption likely occurs at velocities where the absorption is also strong in other species such as H I, C II, and C III.

The O VI absorption is contained within the velocity range of the strong C II λ1036.337 and C III λ977.020 absorption and of the strong H I absorption in the higher Lyman series lines (see Fig. 4). This implies the O VI absorption is kinematically coupled to the absorption produced in the neutral and weakly ionized gas. Such a coupling could occur if the O VI is produced in the interface regions between cool and warm gas and a hotter exterior medium. The interfaces could be of several different types and may involve structures where the behavior of the gas is controlled by conductive heating, radiative cooling, and/or turbulent mixing. See Spitzer (1996) for a discussion of the types of regions where the conditions are favorable for the occurrence of O VI. Conductive interfaces have recently been proposed to explain the O VI absorption produced in Complex C (Fox et al. 2004) and in the thick disk/halo gas along the line of sight to vZ 1128 (Howk et al. 2003).

A lower limit to the number of interfaces along the line of sight to Mrk 421 in the thick Galactic disk can be determined by inspecting the component structure of the gas. The H I 21 cm emission profile reveals four components at –60, –28, –8 and 18 km s$^{-1}$. The more sensitive UV absorption measurements reveal an additional component at –125 km s$^{-1}$ in the higher Lyman series lines and in the C II λ1036.337 line.

With a total of 5 absorbing components between –125 and 70 km s$^{-1}$, there may be 10 possible interface regions (two for each absorption component) between cool and warm neutral and ionized gas and a hotter exterior medium. With a total O VI column density of logN(O VI) = 14.38 associated with the thick disk in this direction over the velocity range from –140 to 60 km s$^{-1}$ the average interface would need to produce an O VI column density of ~2.4x10$^{13}$ cm$^{-2}$ to produce the total observed column density. This number is ~2 times larger than the maximum column density ($1 \times 10^{13}$ cm$^{-2}$) predicted by theories of conductive interfaces (Borkowski, Balbus & Fristrom 1990; Slavin 1989) and ~ 2 times larger than actually measured in the local interstellar medium (Oegerle et al. 2004) where the O VI absorption produced by single clouds can be studied. It therefore appears that conductive interfaces associated with identified ISM components could explain at least 50% of the total O VI absorption associated with the Galactic thick disk. The FUSE resolution is not particularly high for studying the full kinematical complexity of the absorption. Therefore, additional components and interfaces may exist and explain much or all of the remaining O VI absorption. Turbulent mixing in the gas (Slavin, Shull & Begelman 1993) could increase the number of interfaces along the sight line. A modest amount of mixing is expected for the thick disk and halo regions of the Galaxy as they are disturbed by supernovae (see de Avillez & Mac Low 2002). Other processes that might also contribute to the O VI absorption include the cooling gas of a low temperature Galactic Fountain (Houck & Bregman 1990) , the cooling gas of isolated large hot bubbles in the halo created by supernova explosions (Shelton 1998), and the relatively weak O VI absorption that should be directly associated with the O VII and O VIII absorption seen at X-ray wavelengths (see §7).

The intermediate and high velocity clouds along the Mrk 421 line of sight, including IV26 and Complex MII and MIII, are at z < 3.5 kpc. These clouds are probably tracing the cooled return flow of Galactic Fountain gas. Although the O VI absorption associated with IV26 and Complex M may be produced from transition temperature gas in a warm-hot gas conductive interface, the very existence of these structures highlights the strong possibility that some type of flow process is circulating gas from the disk into the halo and back into the disk. Given the complexity of the line of sight to Mrk 421 (see Table 2) it is not surprising that several different O VI production processes are required to explain the thick disk O VI absorption.

Although much of the absorption in the –140 to 60 km s$^{-1}$ velocity range appears to be associated with the thick O VI disk of the Milky Way, we can't rule out additional contributions to the absorption from very distant gas containing O VI. Howk et al. (2002) and Savage et al. (2003) found the O VI absorption associated with the thick disk of the Milky Way to be irregularly distributed over small and large angular scales. For example, toward the LMC and SMC, Howk et al. (2002) found that Milky Way thick disk gas O VI absorption exhibits a peak to peak spread of 0.45 to 0.65 dex in column density over the several degree angular extents of these two galaxies. The small scale irregularities in the distribution of O VI in the thick disk of the Milky Way make it difficult to clearly identify absorption produced by gaseous structures beyond the Milky Way unless the structure introduces enough of a velocity shift to clearly separate the absorption from that associated with the thick disk.

## 6.2. O VI High Positive Velocity Absorption Wing

The nature of the O VI absorption toward Mrk 421 undergoes a sudden change at 60 km s$^{-1}$ (see Figs. 4 and 5). The gas becomes much more highly ionized at $v_{LSR} > 60$ km s$^{-1}$, where the only species detected at UV wavelengths are O VI and C III. The gas in this positive velocity O VI absorption wing toward Mrk 421 is probably related to the anomalous positive velocity wing absorption seen toward the other extragalactic objects in the general direction of Mrk 421 (see Table 1 and Fig. 1).

In the lower panel of Figure 5 we display $N_a(v)$ curves for O VI and C III toward Mrk 421. The vertical line in that panel is placed at $v_{LSR} = 60$ km s$^{-1}$. Over the 60 to 165 km s$^{-1}$ velocity range the O VI and C III absorption profiles have a very similar appearance. In that gas N(O VI)/N(C III) = 10(+4, -6). For the narrower velocity range from 60 to 120 km s$^{-1}$ the ratio is better determined with N(O VI)/N(C III) = 10±3. For the velocity range from 85 to 165 km s$^{-1}$, N(O VI)/N(H I) > 0.17, based on the logN(H I) < 14.19 3$\sigma$ limit provided by the Ly $\gamma$ $\lambda$972.537 line (see Table 5). It is only possible to compare O VI and H I over this more limited velocity range because of strong H I contamination from lower velocity absorption at v < 85 km s$^{-1}$.

It is difficult to determine the physical conditions in the gas producing the O VI absorption wing without having access to other important intermediate and highly ionized species such as Si III, Si IV, and particularly C IV. However, a simple ionization model is clearly ruled out. In gas in collisional ionization equilibrium (CIE; Sutherland & Dopita 1993) at T = 1.8x10$^5$ K with solar abundance ratios for O and C from Allende Prieto, Lambert & Asplund (2001, 2002) N(O VI)/N (C III) = 10 and N(O VI)/ N(H I) = 0.45, consistent with the measurements and limits. However, in such a gas, we would expect to see N(C IV) = 8.9x10$^{13}$ cm$^{-2}$ and N(N V) = 2.3x10$^{14}$ cm$^{-2}$ over the velocity range from 60 to 165 km s$^{-1}$. While we do not have observations extending to wavelengths covering the C IV absorption, we do have a firm 3$\sigma$ upper limit on the N V column density of N(N V) < 10$^{13}$ cm$^{-2}$. Therefore, if the gas in the positive velocity wing has solar elemental abundance ratios, the gas cannot be in CIE at 1.8x10$^5$ K. Such a conclusion is not too surprising since gas in the temperature range (1-5)x10$^5$ K cools very rapidly and is expected to have ionic ratios deviating from CIE values.

The observed O VI to C III ratio and the limits on H I and N V can be achieved in a very low density gas photoionized by the extragalactic background radiation, which we take to be the ionizing background field at z = 0 from AGNs (Madau 1992; Haardt & Madau 1996). Assuming solar abundance ratios for O and C and using the photoionization code CLOUDY (v94.00: Ferland 1996) we can determine the value of the photoionization parameter, U , required to match the observations. U is the ratio of the ionizing photon density to the total gas density. For more details about the photoionization modeling see Sembach et al. (2003). We find the observed value of N(O VI)/N(C III) occurs when log U ∼ −1.2. With this ionization parameter, the total gas density $n(H^o + H^+) = n(H) \sim 5 \times 10^{-6}$ cm$^{-3}$. In such a medium, path lengths of 17 to 195 kpc are required obtain the observed values of N(O VI) and N(C III) for metallicities ranging from solar to 0.1 times solar. This photoionized gas has such a low density that its pressure, P/k, is extremely small. For the solar metallicity case, the model predicts T = $7.6 \times 10^3$ K and P/k = 0.05 cm$^{-3}$ K$^{-1}$ assuming complete ionization of the He. For the 0.1 solar metallicity case, the model predicts T = $1.4 \times 10^4$ K and P/k = 0.1 cm$^{-3}$ K$^{-1}$. Such extremely small pressures would require the gas to be beyond the Milky Way halo in the Local Group medium. The pressure is even substantially smaller than what would seem reasonable for hot gas possibly occupying the medium between the galaxies in the Local Group. In that case we might expect T ∼ $10^6$ K and n(H) ∼ $10^{-5}$ to $10^{-6}$ cm$^{-3}$ and P/k ranging from 23 to 2.3 cm$^{-3}$ K$^{-1}$. Because of the extremely small implied pressure, a photoionization origin of the O VI and C III from the extragalactic background appears unlikely on physical grounds. Given the small number of observed quantities including

N(O VI), N(C III), and limits on N(H I) and N(N V), we have chosen not to consider more complex ionization models.

The kinematic behavior of the positive velocity absorption wing suggests that the absorption might occur in outflowing hot gas, perhaps associated with the turbulent outflow of a Galactic fountain. Such an explanation was proposed for the high positive velocity wing observed in the spectrum of 3C 273 (Sembach et al. 2001). In the case of 3C 273, the wing phenomenon is seen in O VI, but not in C IV, or N V. Unfortunately IGM blending prevents an assessment of the presence of C III. If we are seeing hot outflowing gas along the sight line to Mrk 421, we still have the basic difficulty of explaining why C III and O VI in the wing have the same absorption profile. A possibility is the hot gas outflow contains entrained warm gas which produces the C III profile while the O VI occurs in turbulent or conductive interface regions between the warm gas and the hot ($10^6$ K) gas of the outflow. If we are seeing evidence for outflowing hot gas, the region of the outflow we are sampling must be beyond $z \sim 3.5$ kpc because of the gas distance constraints provided by the two halo stars (see § 4.3).

The most interesting aspect of the O VI wing absorption is the close kinematic coupling of O VI to the C III absorption and the absence of H I. This implies the ionization of the O VI is somehow connected to the presence of the C III. This close kinematic coupling between O VI and C III makes it very doubtful that the O VI in the high positive velocity wing toward Mrk 421 co-exists with the much hotter gas required to explain the X-ray observations of O VII and O VIII absorption (see the next section).

## 7.   O VI, O VII  and O VIII Absorption near z = 0.0.

Strong  O VII and O VIII  absorption at zero redshift has been found with XMM-Newton (Rasmussen et al. 2003) and Chandra (Nicastro 2004).  Rasmussen et al. report O VII Heα λ21.602  and O VIII Lyα λ18.970 equivalent widths of 15.4±1.7 mÅ and 4.3±1.1 mÅ, respectively. These correspond to O VII and O VIII velocity equivalent widths of 214 and 68 km s$^{-1}$,  respectively.  In contrast the total velocity equivalent width of the O VI λ1031.926 integrated over velocity range from –140 to 165 km s$^{-1}$ is only 84 km s$^{-1}$.

A lower limit to the column densities of O VII and O VIII can be obtained by assuming the lines lie on the linear part of the curve of growth.  Adopting the f values from Verner et al. (1996), the O VII and O VIII column densities are 5.4x10$^{15}$ and 3.2x10$^{15}$ cm$^{-2}$, respectively.  However, line saturation will cause the true values to be larger.  Rasmussen et al. (2003) have constrained the value of b for O VII to lie between 200 and 560 km s$^{-1}$ based on the observed width of the O VII Heα line.  With this constraint, the value of N(O VII) lies between 5.4x10$^{15}$ and 7.2x10$^{15}$ cm$^{-2}$.  For a plot of the curve of growth for the O VII line with observational results for three extragalactic lines of sight including Mrk 421 see Figure 2a in Futamoto et al. (2004).  In our discussions below we will adopt the value 5.4x10$^{15}$ cm$^{-2}$ for N(O VII) but note that more reliable values of the  column densities  along the line of sight to Mrk 421 will become available when the high signal-to-noise Chandra X-ray observations (Nicastro 2004) are published.

In Table 9 we list values of N(O VII)/N(O VIII), N(O VI)/N(O VII) for a plasma in collisional ionization equilibrium (Sutherland & Dopita 1993) with values of log T ranging from 5.5 to 7.0.   It is of interest to determine the column density of O VI that might be associated with the observed column density of O VII.   Assuming N(O VII) = $5.4 \times 10^{15}$ cm$^{-2}$ we have listed in column 4 of Table 9 the expected column density of O VI at the different listed temperatures.  We similarly list the expected column density of H I assuming a solar abundance ratio for (O/H).  The observed upper limit of logN(H I) < 13.57  for v = 110 to 165 km s$^{-1}$ is compatible with  5.7 < log T  < 6.4.   For logT  > 6.0 the H I line width will be so large  ($b$ = 129 km s$^{-1}$),  it would be extremely difficult to see the small amount of H I associated with the hot gas.  If the O VII exists in gas with logT = 6.0 to 6.5,  a detectable amount of O VI should co-exist with the O VII.  For log T= 6.0 and 6.5,  N(O VI) is expected to be $2.1 \times 10^{13}$  and  $2.8 \times 10^{13}$ cm$^{-2}$.  For comparison, the column density of O VI in the high positive velocity wing from 60 to 165 km s$^{-1}$ is $4.6 \times 10^{13}$ cm$^{-2}$.  However, we have already established that it is unlikely that the O VI in this wing co-exists with the O VII because at the high temperatures required to produce O VII we would not expect to detect C III.   We must therefore look at other velocities to find evidence for O VI absorption that might be associated with the O VII absorption.

There is no evidence for O VI absorption between –1000 to –140 km s$^{-1}$ and 165 to 1200 km s$^{-1}$.  Therefore, any O VI absorption associated with the O VII must occur in the velocity range from –165 to 60 km s$^{-1}$ or else have such a large value of $b$ that it can not be discerned against the Mrk 421 continuum at the larger negative or positive velocities. If the $b$ value for the associated O VI absorption were as large as 200 km s$^{-1}$ (FWHM = 333 km s$^{-1}$) with a column density of $(2.1\text{-}2.8\text{-}) \times 10^{13}$ cm$^{-2}$ the line would only be 2-3% deep and extremely difficult to detect.  If the associated O VI coincides with the strong thick disk absorption over the velocity range –140 to 60 km s$^{-1}$,  the small expected column density would be difficult to separate from the much stronger thick disk absorption.   This exercise revels that it is difficult to convincingly separate absorption produced by O VI co-existing with the O VII from  the various other possible sites of O VI absorption in the Galaxy along the line of sight to Mrk 421.

If there is Local Group hot gas along the line of sight to Mrk 421, we might expect it to roughly follow the velocity pattern of Local Group galaxies.  The six Local Group galaxies within ~40 degrees of  Mrk 421 have LSR radial velocities from  ~-84 to 303 km s$^{-1}$ with  four of the six between 218 and 303 km s$^{-1}$.  Since there is no evidence for O VI with v > 160 km s$^{-1}$ this would suggest that hot Local Group gas containing detectable O VI probably does not exist in the general direction of Mrk 421.

The strong O VII and O VIII absorption could be produced by a highly extended (~100 kpc)  corona to the Milky Way, a hot phase of Local Group gas, or some  or a combination of the two.   A  better understanding of the strength of the Milky Way coronal contribution will be the key to determining the relative contributions from the two sites of orgin.  Such an understanding might come from measures of X-ray absorption toward sources situated far from the disk of the Galaxy but within the extent of corona.   Factors of  five to ten improvements in the X-ray resolution from  the current 750-900 km s$^{-1}$ to  ~100-200  km s$^{-1}$ might provide the enhanced kinematic information required to separate Milky Way corona absorption from Local Group hot gas absorption.

Possible evidence for a very strong Galactic ISM  contribution to O VII and O VIII absorption seen toward extragalactic objects comes from the detection by Futamoto et al. (2004) of strong O VII, O VIII, and Ne IX absorption toward the low mass X-ray binary 4U1820-303 in the globular star cluster NGC 6624 at 7.6±0.4 kpc in the direction $l = 2.8^{\circ}$ and $b = -7.9º$.   However, this low latitude line of sight passes through a very active region of the Milky Way. The measurements could be dominated by gas near the Galactic Center.  Therefore, it will be important to extend such observations of O VII and O VIII to other distant Galactic X-ray sources away from the Galactic Center, preferably at high Galactic latitudes.

## 8.   Summary

We discuss new high quality FUSE far-UV spectroscopic observations covering the wavelengths 910 to 1187 Å of ISM and IGM absorption toward the X-ray bright AGN Mrk 421 obtained at a resolution of 20 km s$^{-1}$.  The extragalactic measurements are supplemented with FUSE  observations of the distant halo stars BD +38  2182 and HD 93521 in order to obtain information on the distances to the various absorbing structures seen toward Mrk 421.  The emphasis of the investigation is on the origin of the O VI absorption toward Mrk 421 and the possible relationships among the absorption produced by O VI  and the X-ray absorption by O VII and O VIII near zero redshift observed by XMM  and Chandra.  Our principal results can be summarized as follows:

1. The line of sight to Mrk 421 in the direction $l = 179.83$  and $b = +65.15$  samples many different ISM structures, including the Local Bubble, an IVC in the low halo, the thick disk of  warm ionized gas,  the very thick disk of hot ionized gas, and the gas in the outer boundaries of HVC Complex M.   In addition, a  high positive velocity wing of O VI absorption extending to 165 km s$^{-1}$ may be tracing the outflow of hot gas  into the Galactic Halo.  Beyond the Local Group of galaxies, the line of sight passes through a Ly $\alpha$ absorber at $cz = 3022$ km s$^{-1}$ that is situated in a galactic void.

2.  The new FUSE observations of Mrk 421 provide high quality measures of O VI $\lambda\lambda$1031.93, 1037.62 absorption and the absorption produced by various cooler ISM species,  including H I, C II, C III, O I, N I, N II, Fe II, Fe III, and S III.

3. The O VI absorption extending  from –140 to 60 km s$^{-1}$ is associated with strong absorption from warm and cool gas.  Much of  the O VI absorption in this velocity range likely arises in the Galactic thick disk, which has an  extension away from the Galactic plane described by a layer with a scale height of ~ 4 kpc in the direction to Mrk 421.  The ionization of the O VI in the thick disk  is likely from a combination of processes, with ~50% of the O VI absorption occurring in conductive interfaces between cool and warm gas and a hot ISM.  The remainder of the O VI absorption could be from cooling Galactic fountain gas or hot gas bubbles in the halo. Extragalactic contributions to the O VI absorption over this velocity range can not be ruled out.

4. The O VI absorption extending from 60 to 165 km s$^{-1}$ has very little associated lower ionization absorption.  The only other species detected is C III.   Over the velocity range from 60 to 165 km s$^{-1}$ the O VI and C III profiles are very similar, with N(O VI)/N(C III) =10±2 from 60 to 120 km s$^{-1}$.  The positive velocity wing of O VI absorption appears to trace gas in the Galactic halo or beyond, since the high positive velocity wing is not detected toward the stars BD +38  2182 at z = 3.5 kpc or HD 93521 at z = 1.8 kpc. The  C III and O VI absorption might occur in outflowing hot gas perhaps associated with the outflow of a Galactic fountain.  The close association of O VI and C III implies the O

VI absorption over the 60 to 165 km s$^{-1}$ velocity range does not likely co-exist with the strong O VII and O VIII absorption detected by Chandra and XMM.

6. The O VII and O VIII absorption detected by Chandra and XMM may trace the hot gas in a highly extended Galactic Corona or trace the superposition of the absorption by gas in a Galactic Corona and a Local Group medium. Unfortunately, the low spectral resolution of the existing X-ray observations makes it difficult to distinguish between these two possibilities. Suggestions by Nicastro et al. (2002) and Nicastro (2004) that the signature of a Local Group WHIM filament has definitely been detected in high velocity O VI not only fail to account for the co-existence of O VI and C III, but are also difficult to reconcile with the absence of O VI absorption in the velocity range 200-400 km s$^{-1}$ expected for Local Group galaxies in the direction of Mrk 421.

7. The Lyman $\alpha$ absorber at $z = 0.01$ or $cz = 3022$ km s$^{-1}$ with logN(H I) = 13.25±0.04 occurs in a galactic void. This absorber is not detected in the lines of C II, C III, N V, or O VI. No O VI is detected near the redshift where Nicastro (2004) has reported the detection of O VII.

We thank George Sonneborn and the FUSE observational team for their help in obtaining the target of opportunity FUSE observations of Mrk 421 in early 2003. This work is based on data obtained by the NASA-CNES-CSA FUSE mission operated by the Johns Hopkins University. Thanks are extended to the FUSE development and operational teams for producing and operating such a capable far-UV observatory. BDS acknowledges financial support from NASA grant NNG04GC70G. BPW acknowledges support from NASA grants NAG5-9024 and NAG5-9179. KRS acknowledges support

TABLE 1
O VI ABSORPTION LINE OBSERVATIONS FOR OBJECTS NEAR THE
DIRECTION TO MRK 421

| Object | Type | l | b | z/v(km s⁻¹) | $z^a$ (kpc) | logN(O VI) thick disk[b] (\|v\| <~100 km s⁻¹) | logN(O VI) +velocity wing[b] (v >~100 km s⁻¹) | Ref. |
|---|---|---|---|---|---|---|---|---|
| Mrk 421 | BL Lac | 179.83 | 65.03 | 0.03 | … | 14.38±0.02 | 13.66±0.05 | 1 |
| Mrk 36 | Gal | 201.76 | 66.49 | 614 | … | 14.36±0.11 | No wing | 2,3 |
| HS1102+3441 | QSO | 188.56 | 66.22 | 0.510 | … | 14.71±0.07 | 14.30±0.10 | 2,3 |
| NGC 3504 | Sey | 204.60 | 66.04 | 1534 | … | <14.19 (3σ) | No Wing | 2,3 |
| Mrk 162 | Gal | 166.26 | 62.41 | 6458 | … | 14.63±0.04 | No wing | 7 |
| Mrk 153 | Gal | 156.73 | 56.01 | 2565 | … | 14.26±0.17 | No wing | 7 |
| Ton 1187 | QSO | 188.33 | 55.38 | 0.0700 | … | 14.35±0.08 | No wing | 2,3 |
| NGC 3310 | Sey | 156.60 | 54.06 | 993 | … | 14.56±0.02 | No wing[c] | 2, 3 |
| NGC 3353 | Sey | 152.30 | 53.37 | 944 | … | 14.62±0.05 | No wing | 7 |
| PG 1001+291 | QSO | 200.08 | 53.21 | 0.3297 | … | 14.50±0.06 | 13.97±0.13 | 2,3 |
| Mrk 33 | Gal | 152.20 | 52.80 | 1461 | … | 14.67±0.07 | No wing | 7 |
| PG 0953+414 | QSO | 179.79 | 51.71 | 0.234 | … | 14.45±0.03 | 13.83±0.06 | 2,3 |
| PG 0947+396 | QSO | 182.85 | 50.75 | 0.2060 | … | 14.54±0.07 | 14.19±0.10 | 2,3 |
| BD +38 2182 | B3 V | 182.16 | 62.21 | 80 | 3.5 | 14.17±0.03 | No wing | 4,6 |
| HD 93521 | O9.5 V | 183.14 | 62.15 | −16 | 1.8 | 14.01±0.02 | No wing | 5,6 |

[a] The distance away from the Galactic plane is listed for the two stellar sources.

[b] The O VI column densities listed refer to O VI associated with the thick disk and in the positive velocity wing. The thick disk column densities refer to O VI absorption velocities |v| < ~ 100 km s⁻¹ (see Wakker et al. 2003 for exact velocity ranges). The positive velocity wing column densities typically represent integrations over the velocity range from +100 to ~+250 km s⁻¹ (see Wakker et al. 2003 and Sembach et al. 2003). The wing absorption may occur in outflowing Galactic Fountain gas. However, origins in gas beyond the Milky way are also possible.

[c] The continuum in the velocity range of the positive velocity O VI absorption wing is uncertain for NGC 3310.

TABLE 2
ISM AND IGM STRUCTURES ALONG THE MRK 421 LINE OF SIGHT

| Structure | $v_{LSR}$ (km s$^{-1}$) | d (kpc) | Comment |
|---|---|---|---|
| Local Bubble....... | ~0 | <0.2 | The Local Bubble appears to open into the halo (Welsh et al. 1999). |
| H I clouds....... | −8 | 0.2 to 0.4 | This neutral cloud is probably beyond the boundaries of the Local Bubble. |
| IV26....... | −60 | 0.4 to 1.8 | The distance limit is from Wakker (2001). The near solar abundances in the IVCs in this direction suggest they trace the return flow of a Galactic Fountain. |
| Cloud MI....... | −120 | ? | The structure edge is ~4 east of Mrk 421. |
| Cloud MII....... | −90 | <3.5 | The structure edge is ~2 west of Mrk 421. Distance limit is from Danly et al. (1993). |
| Cloud MIII....... | −90 | <3.5 | The structure edge is ~4 south of Mrk 421. Distance limit is from Danly et al. (1993). |
| WIM....... | −100 to 60 | 1 | One scale height is given for the distance of the gas in the WIM. |
| HIM associated with the thick disk O VI... | −100 to 60 | 4 | One scale height is given for the distance of the thick disk O VI. |
| O VI high positive velocity wing...... | 60 to 165 | > 3.5 | Distance limit is based on the absence of the O VI wing toward the stars HD +38 2182 and HD 93521. |
| Extended hot corona or hot Local Group gas....... | ? | > 70 kpc | Zero redshift O VII and O VIII absorption suggests that either a very extended hot Galactic Corona or hot Local Group gas is present in the direction of Mrk 421. |
| Lyα cloud....... | 3022 | 42 Mpc | H I cloud in a Galactic Void |

TABLE 3
KNOWN GALAXIES BEYOND THE LOCAL GROUP NEAR
THE MRK 421 LINE OF SIGHT[a]

| Galaxy | $l$ (°) | $b$ (°) | type | V (mag.) | v (km s$^{-1}$) | Dist (Mpc) | D[b] (kpc) | ρ[c] (Mpc) |
|---|---|---|---|---|---|---|---|---|
| IRAS11011+3830 | 179.84 | 64.93 | … | 16.39 | 9241 | 135.7 | 22.1 | 0.24 |
| 2MASXJ11050808+3807525 | 179.91 | 65.18 | … | 16.18 | 9350 | 137.2 | 18.8 | 0.37 |
| CGCG213-027 | 179.64 | 65.19 | … | 15.4 | 8465 | 124.6 | 14.5 | 0.39 |
| CGCG184-053 | 180.04 | 65.2 | … | 15.4 | 8740 | 128.5 | 22.4 | 0.44 |
| CGCG184-040NED02 | 185.05 | 64.27 | … | 16.5 | 665 | 12.7 | 0.4 | 0.52 |
| SBS1054+365 | 185.06 | 64.27 | … | 16 | 603 | 12.7 | 0.4 | 0.52 |
| CG803 | 183.58 | 64.77 | … | 18 | 1283 | 22.0 | 0.6 | 0.62 |
| CGCG184-054 | 180.12 | 65.27 | … | 15.5 | 9020 | 132.5 | 22 | 0.62 |
| NGC3432 | 184.77 | 63.16 | SB(s)m | 11.67 | 616 | 12.4 | 24.6 | 0.62 |
| NGC3600 | 170.29 | 65.68 | Sa? | 12.6 | 719 | 13.9 | 16.6 | 0.98 |
| HS1059+3934 | 177.74 | 64.25 | HII | 17.7 | 3274 | 50.4 | 3.8 | 1.05 |
| IC2620 | 179.47 | 64.56 | - | 14.9 | 8586 | 126.3 | 40.4 | 1.10 |
| UGC6817 | 166.20 | 72.75 | Im | 13.4 | 242 | 7.3 | 8.7 | 1.15 |
| IC2615 | 180.76 | 64.66 | - | 15.5 | 8481 | 124.8 | 18.2 | 1.18 |
| UGC5829 | 190.07 | 61.53 | Im | 13.73 | 629 | 12.6 | 17.3 | 1.27 |
| NGC3413 | 193.35 | 63.48 | S0 | 13.08 | 645 | 12.9 | 8.3 | 1.37 |
| UGC6161 | 167.86 | 63.19 | SBdm | 14 | 756 | 14.3 | 10.8 | 1.38 |
| NGC3319 | 175.98 | 59.34 | SB(rs)cd | 11.48 | 739 | 14.0 | 25.2 | 1.45 |
| UGC5996 | 176.83 | 62.22 | Sdm | 16.5 | 1621 | 26.7 | 8.5 | 1.45 |
| UGC6307 | 177.96 | 67.52 | Sdm: | 14.78 | 1964 | 31.8 | 12.9 | 1.45 |
| | | | | | | | | |
| HS1059+3934 | 177.74 | 64.25 | HII | 17.7 | 3274 | 50.4 | 3.8 | 1.05 |
| KUG1102+413 | 173.40 | 64.2 | … | 17.66 | 2988 | 46.3 | 2.7 | 2.32 |
| KUG1122+359 | 182.78 | 69.75 | S | 15.3 | 2570 | 40.5 | 7.1 | 3.43 |
| UGC6491 | 184.07 | 70.84 | Sdm | 15 | 2530 | 40.0 | 17.5 | 4.2 |
| UGC6149 | 168.77 | 63.26 | S | 14.6 | 3208 | 49.3 | 14.4 | 4.42 |
| UGC6383 | 166.26 | 66.02 | Sd | 15.34 | 3156 | 48.7 | 17.0 | 4.84 |
| NGC3415 | 170.53 | 60.76 | SA0+: | 13.45 | 3303 | 50.6 | 30.9 | 5.3 |
| KUG1130+337 | 187.39 | 72.05 | S | 15.4 | 2574 | 40.7 | 10.7 | 5.35 |
| NGC3416 | 170.42 | 60.76 | S? | 15.4 | 3338 | 51.1 | 8.9 | 5.39 |
| CGCG186-009 | 188.26 | 72.33 | S0? | 15.02 | 2583 | 40.8 | 5.9 | 5.63 |
| CGCG242-008 | 163.78 | 63.23 | … | 15.3 | 2949 | 45.6 | 6.6 | 5.74 |
| UGC5953 | 169.01 | 60.33 | Pec | 13.8 | 3191 | 49.0 | 8.5 | 5.83 |
| UGC6545 | 190.11 | 72.31 | Sb | 14.7 | 2619 | 41.4 | 12 | 5.89 |
| HS1135+3709 | 176.1 | 71.73 | … | 18.1 | 3251 | 50.3 | 1.5 | 6.0 |

[a] The top half of this table lists galaxies close to the line of sight to Mrk 421 from the NASA Extragalactic Database (NED) with impact parameters, ρ < 1.5 Mpc, regardless of velocity. The lower half of this table gives galaxies along the line of sight with ρ < 6 Mpc and velocity in the range from 2500 and 3500 km s$^{-1}$. The absorber at z = 0.01 or cz = 3022 km s$^{-1}$ lies in a region with no known galaxies near the redshift and within 200 kpc to a limit $M_B$ = −14.5 (McLin et al. 2002). The galaxy positions with respect to Mrk 421 are shown in Figure 2.

[b] Physical size of the galaxy.

[c] Impact parameter between Mrk 421 and the galaxy.

TABLE 4
FUSE OBSERVATIONS

| object | Date (yr.mn.dy) | Program ID | Obs. ID | Aperture | $t_{actual}$ (ks) |
|---|---|---|---|---|---|
| Mrk 421 | 2000.12.01 | P101 | 2901 | LWRS | 21.6 |
| Mrk 421 | 2003.01.19 | Z010 | 0101 | LWRS | 26.5 |
| Mrk 421 | 2003.01.20 | Z010 | 0102 | LWRS | 23.4 |
| Mrk 421 | 2003.01.21 | Z010 | 0103 | LWRS | 12.1 |
| HD 93521 | 2002.11.25 | P101 | 2603 | HIRS | 7.3 |
| HD 93521 | 2003.05.20 | P101 | 2604 | MDRS | 7.6 |
| BD +38  2182 | 2000.11.30 | P101 | 2801 | LWRS | 12.0 |

TABLE 5
SELECTED ABSORPTION LINES IN THE FAR-UV SPECTRUM OF MRK 421

| Species | $\lambda^a$ (Å) | $\log(f\lambda)^a$ | Velocity Range$^b$ (km s$^{-1}$) | $W_\lambda \pm \sigma$ (mÅ) | $\log N \pm \sigma$ (N in cm$^{-2}$) | Notes |
|---|---|---|---|---|---|---|
| O VI…... | 1031.926 | 2.138 | −140 to 60 | 237±9 | 14.38±0.02 | 1 |
| O VI…... | 1037.617 | 1.836 | −140 to 60 | 150±10 | 14.44±0.04 | 1,3 |
| N V…..... | 1238.821 | 2.289 | −140 to 60 | <36 (3σ) | <13.23 (3σ) | 2 |
| | | | | | | |
| O VI…... | 1031.926 | 2.138 | 60 to 165 | 52±9 | 13.66±0.06 | 1 |
| O VI…... | 1037.617 | 1.836 | 60 to 165 | 17±8 | 13.46(+0.16,-0.27) | 1,3 |
| N V…..... | 1238.821 | 2.289 | 60 to 165 | <26 (3σ) | < 13.03 (3σ) | 2 |
| N V…..... | 1241.804 | 1.988 | 60 to 165 | <26 (3σ) | < 13.33 (3σ) | 2 |
| C III…... | 977.020 | 2.875 | 60 to 165 | 26±10 | 12.66 (+0.13,-0.19) | 1 |
| C II…..... | 1036.337 | 2.102 | 60 to 165 | … | … | 4 |
| N II…..... | 1083.994 | 2.072 | 60 to 165 | <56 (3σ) | <13.75 (3σ) | 2 |
| Fe III…... | 1122.524 | 1.952 | 60 to 165 | <20 (3σ) | <13.37 (3σ) | 2 |
| | | | | | | |
| O VI…... | 1031.926 | 2.138 | 85 to 165 | 31±9 | 13.42±0.08 | 1 |
| N V…..... | 1238.821 | 2.289 | 85 to 165 | <23 (3σ) | < 13.04 (3σ) | 2 |
| C III…... | 977.020 | 2.875 | 85 to 165 | <25 (3σ) | < 12.61 (3σ) | 1,5 |
| H I Ly β... | 1025.722 | 1.909 | 110 to 165 | <30 (3σ) | <13.61 (3σ) | 2,6,7 |
| H I Ly γ... | 972.537 | 1.450 | 85 to 165 | <38 (3σ) | <14.19 (3σ) | 2,7 |
| H I Ly δ... | 949.743 | 1.122 | 85 to 165 | <38 (3σ) | <14.53 (3σ) | 2,7 |
| H I Ly ε... | 937.804 | 0.864 | 85 to 165 | <38 (3σ) | <14.80 (3σ) | 2,7 |
| H I Ly ζ... | 930.748 | 0.652 | 85 to 165 | <38 (3σ) | <15.01 (3σ) | 2,7 |
| H I Ly η... | 926.226 | 0.471 | 85 to 165 | <40 (3σ) | <15.22 (3σ) | 2,7 |

$^a$ Rest wavelengths and values of $\log(f\lambda)$ are from Morton (2003).

$^b$ The velocity ranges for the equivalent width and column density integrations are listed. For the high positive velocity wing we list ranges from 60 to 165 km s$^{-1}$ and for the more restricted ranges from 85 to 165 km s$^{-1}$ and from 110 to 165 km s$^{-1}$. Because of blending of H I absorption from low velocity gas with absorption at high velocity, we are only able to place limits on the H I high velocity absorption over the more limited ranges in velocity.

Notes: (1) Column density from the apparent optical depth method of Savage & Sembach (1991). (2) The 3σ column density limit is derived from the equivalent width limit and assumes no line saturation. (3) Contamination from H$_2$ R(1) 1038.156 has been removed. (4) C II at these velocities is contaminated by C II*λ1037.018. (5) The formal equivalent width measurement for C III is $W_\lambda$ = 12±8. The 3σ limit is listed. (6) Ly β absorption is contaminated by terrestrial emission for v <110 km s$^{-1}$. (7) Due to the presence of strong H I absorption at lower velocity, it is only possible to compare high velocity O VI and H I at v > 85 km s$^{-1}$.

TABLE 6
H I COMPONENT FITS TO THE LYMAN SERIES ABSORPTION OF MRK 421[a]

| $v_{LSR}$ (km s$^{-1}$) | b (km s$^{-1}$) | N(H I) ($10^{18}$ cm$^{-2}$) |
|---|---|---|
| −125 | 19±3 | 4.8 (+6.9, -2.8) |
| −60 | 25(+6, -5) | 27 (+9, -7) |
| −8 | 28 (+7, -6) | 75 (+29, -17) |

[a] The HI 21 cm emission observations (Fig. 2a) and FUSE O I absorption profiles (Fig. 4) were used to partly constrain the H I Lyman series absorption line fit results listed here. The H I 21 cm components at −28 and +18 km s$^{-1}$ seen in Fig.2a are subsumed into the −8 km s$^{-1}$ component required to fit the Lyman series absorption listed above. The fit to the Lyman series reveals the existence of the H I component at −125 km s$^{-1}$ which is also seen in C II λ1036.337 absorption.

TABLE 7
O VI λ1031.926 ABSORPTION TOWARD MRK 421, BD +38 2182, and HD 93521

| Object | Velocity Range (km s$^{-1}$) | $W_\lambda \pm \sigma_{sc}$ (mÅ) | logN(O VI)±σ (N in cm$^{-2}$) | log[N(O VI)sin\|b\|]±σ (N in cm$^{-2}$) |
|---|---|---|---|---|
| Mrk 421……….. | −140 to 165 | 287±11 | 14.46±0.02 | 14.42±0.02 |
| "…………. | −140 to 60 | 237±9 | 14.38±0.02 | 14.34±0.02 |
| "…………. | 60 to165 | 52±9 | 13.66±0.06 | 13.62±0.06 |
| "………….. | −100 to 60 | 214±10 | 14.34±0.03 | 14.30±0.03 |
| BD +38 2182… | −100 to 60 | 142±10 | 14.17±0.03 | 14.12±0.03 |
| "…………. | 60 to 165 | <21(3σ) | <13.22(3σ) | <13.17(3σ) |
| HD 93521……….. | −100 to 60 | 108±5 | 14.01±0.02 | 13.96±0.0 |
| "…………. | 60 to 165 | <24(3σ) | <13.28(3σ) | <13.23(3σ) |

TABLE 8
ABSORPTION LINES IN THE SPECTRUM OF MRK 421 AT  cz = 3022 km s$^{-1}$

| Species | $\lambda_{rest}$[a] (Å) | log(f$\lambda$)[a] | $W_\lambda$(rest)$\pm\sigma$ (mÅ) | logN$\pm\sigma$ (N in cm$^{-2}$) | Note |
|---|---|---|---|---|---|
| H I L$\alpha$ | 1215.670 | 2.704 | 82$\pm$8 | 13.25$\pm$0.04 | 1 |
| H I L$\beta$ | 1025.722 | 1.909 | … | … | 2 |
| H I L$\gamma$ | 972.537 | 1.450 | <34 (3$\sigma$) | <14.15 (3$\sigma$) | 3 |
| O VI | 1031.926 | 2.137 | <19(3$\sigma$) | <13.18 (3$\sigma$) | 3, 4 |
| O VI | 1037.617 | 1.836 | … | … | 5 |
| N V | 1238.821 | 2.289 | < 22 (3$\sigma$) | <13.02 (3$\sigma$) | 3 |
| N V | 1242.804 | 1.988 | < 19 (3$\sigma$) | <13.26 (3$\sigma$) | 3 |
| C III | 977.020 | 2.872 | <33 (3$\sigma$) | <12.72(3$\sigma$) | 6 |
| C II | 1036.337 | 2.102 | <30(3$\sigma$) | <13.43(3$\sigma$) | 3 |

[a] Rest wavelengths and values of log($\lambda$f) are from Morton (2003).

Notes: (1)  The velocity of Ly $\alpha$  is 3022$\pm$2 km s$^{-1}$ based on our analysis. The value of logN(H I)  is from the apparent optical depth method.  The value of b  determined from  the observed line profile corrected for instrumental blurring is 29.5$\pm$5.6  km s$^{-1}$. (2) Ly $\beta$ $\lambda$1025.722 is strongly blended with ISM C II $\lambda$1036.337.  (3) The 3$\sigma$ limits  to W$_\lambda$ listed are based on a quadrature sum of the statistical error, the continuum placement error, and an estimate to  the fixed pattern noise error of 5mÅ.  The corresponding limits to logN assume the absorption is on the linear part of the COG. The metal line equivalent width and column density limits refer to integrations over the velocity range from 2990 to 3050 km s$^{-1}$.  (4) The O VI $\lambda$1031.926 limit at 3022 km s$^{-1}$ is from the LiF1A segment observations. (5) The O VI 1037.617 line blends with ISM Ar I $\lambda$1048.220.  A useful limit for absorption at 3022 km s$^{-1}$ can't be obtained. (6) The C III $\lambda$977.020 line expected at 3022 km s$^{-1}$  falls in the positive velocity wing of the relatively strong ISM H$_2$ R(1) $\lambda$986.764  line.  We can only report a 3$\sigma$  limit for W$_\lambda$ and logN.  The W$_\lambda$ limit includes the effects of the deblending uncertainty.  The logN limit assumes the line is on the linear part of the COG.

TABLE 9

IONIC RATIOS IN CIE AND O VI ABSORPTION POSSIBLY ASSOCIATED WITH
O VII AND O VII[a]

| logT(K) | O VII/O VIII | O VI /O VII | logN(O VI)[b] (cm$^{-2}$) | logN(H I)[c] (cm$^{-2}$) | b(H I)[d] (km s$^{-1}$) |
|---|---|---|---|---|---|
| 5.5 | … | 0.28 | 15.18 | 14.35 | 72 |
| 5.75 | 7600 | 0.017 | 13.95 | 13.78 | 98 |
| 6.0 | 54 | 0.0039 | 13.33 | 13.46 | 129 |
| 6.5 | 4.7 | 0.0052 | 13.44 | 13.97 | 229 |
| 7.0 | 0.019 | 0.0022 | 13.07 | … | … |

[a] Ionic ratios for O VII/O VIII and O VI/O VII are listed for a hot plasma in collisional

ionization equilibrium from Sutherland and Dopita (1993). O VI, O VII, and O VIII peak in abundance at log T = 5.45, 5.90, and 6.35 respectively.

[b] Expected column density of O VI if collisional ionization equilibrium exists in the gas

responsible for O VII absorption toward Mrk 421 with N(O VII) = $5.4 \times 10^{15}$cm$^{-2}$. This

estimate for N(O VII) uses the O VII equivalent width from Rasmussen et al. (2003) and

assumes little or no line saturation, which implies $b > 500$ km s$^{-1}$. The true O VII column

density could be larger. If O VII is mostly formed in gas with $10^7 > T > 10^6$ K, the O VI

column density in the hot gas is expected to range from $\sim 1 \times 10^{13}$ to $3 \times 10^{13}$ cm$^{-2}$, which represents at most ~10% of the O VI detected along the Galactic line of sight to Mrk 421.

[c] The expected column density of H I if CIE exists in the gas responsible for O VII

absorption toward Mrk 421 with N(O VII) = $5.4 \times 10^{15}$ cm$^{-2}$. The value of N(H I) assumes

O/H in the hot gas is 0.1 times the solar value of Allende Prieto et al. (2001).

[d] The thermal Doppler parameter for possible associated H I absorption lines.

FIG. 1. - 21 cm observations from the Leiden–Dwingeloo survey of intermediate and high velocity H I in the general direction of Mrk 421 are displayed as a function of Galactic longitude and latitude. The grey scale and contours show H I emission with $v_{LSR}$ = –200 to –90 km s$^{-1}$ [panel a; contours at N(H I)= 5, 15, 25, 50, 75x10$^{18}$ cm$^{-2}$ ]; $v_{LSR}$ = –90 to –30 km s$^{-1}$ [panel b; contours at N(H I)= 5, 50, 100, 200 x10$^{18}$ cm$^{-2}$ ] ; $v_{LSR}$= 30 to 100 km s$^{-1}$ [panel c; contours at N(H I) = 5, 15, 25, 50x10$^{18}$ cm$^{-2}$]; and $v_{LSR}$ = 100 to 200 km s$^{-1}$ [panel d; contours at N(H I)= 5, 15, 25, 50x10$^{18}$ cm$^{-2}$]. The symbols identify objects for which O VI observations with S/N at O VI >3 in the region $l$ = 150 to 210 and $b$ = 45 to 75 have been obtained (see Table 1). Open circles are for extragalactic directions where the large positive velocity O VI absorption wings have been detected with circle size proportional to N(O VI) in the wing. The downward pointing triangles are for extragalactic and galactic directions with no O VI positive velocity wing with the triangle size proportional to the O VI column density limit in the positive velocity wing.

FIG. 2. - Selected UV absorption lines in the FUSE spectra of (a) Mrk 421, (b) BD +38 2182, and (c) HD 93521. From top to bottom in each figure we illustrate in the different panels: (1) Flux versus wavelength in Å for a wide wavelength range centered on 1032Å showing the general behavior of the continuum. (2) - (3) Line profile plots (flux versus LSR velocity) for O VI λ1031.926 and O VI λ1037.617. (4) N$_a$(v) versus LSR velocity for the O VI λ1031.926 (heavy) and O VI λ1037.617 (light) lines. (5)-(7) Line profile plots for C II λ1036.337, Si II λ1020.699, and Ar I λ1048.220. (8)-(9) Brightness temperature versus LSR velocity for 21 cm H I emission observations centered on or near

the object. For Mrk 421 the continua including the effects of $H_2$ absorption are shown for all the absorption line profiles on the various panels.  For the two stars the continua and $H_2$ absorption are only shown for the O VI $\lambda$1031.926 absorption line region of the spectrum.  For Mrk 421 the radio observations are from the Green Bank 43 m telescope. For BD +38  2182 and HD 93521 the radio observations are from 76 m Jodrell Bank telescope.   The velocities, amplitudes, FWHM, and value of N(H I) in the different H I 21 cm emission components are listed at the bottom of the figure.  Note that these three figures have the same format as those illustrated for 100 AGNs in Wakker et al. (2003). The positions of contaminating $H_2$ absorption lines are indicated with the branch (R or P) and J value  (1, 2, 3, or 4).  The far-UV fluxes are plotted in the units $10^{-14}$ erg cm$^{-2}$ s$^{-1}$ Å$^{-1}$.

FIG. 3.- Galaxies and galaxy groups in the region  l = 165  - 195  and b = 58  - 74  with v < $10^4$ km s$^{-1}$.  The galaxies (see Table 3) are from NED, with the symbol shape coded according to cz and the symbol size coded according to the physical size of the galaxy. The five circles outline the boundaries of galaxy groups from the catalog of Geller & Huchra  (1983).  Galaxies in a nearby grouping  (the Leo Spur group, Tully 1988) with v = 620±160 km s$^{-1}$ cover the entire field shown in the figure.   The horizontal bars in the legend on the top right indicate the distances on the plot corresponding to 1 Mpc at a distance  implied by the mid-point of the range in cz.   Mrk 421 lies in a direction relatively devoid of galaxies,  except for the range  cz = 8500-9500 km s$^{-1}$.

FIG. 4 -   Flux ($10^{-14}$ erg cm$^{-2}$ s$^{-1}$ Å$^{-1}$) versus LSR velocity for the absorption toward Mrk 421 for a wide range of  ions as marked on each of the panels in Figs. 4a, 4b, and 4c .

The vertical dotted lines appear at $v_{LSR}$ = 60, and 165 km s$^{-1}$. Each of the continua shown was fitted in a small region near each individual absorption line. The same continuum was used for closely spaced absorption lines. The model fit to the H I Lyman lines is discussed in §4.2. The positions of contaminating ISM lines are marked in each panel. The numbers 1, 2, 3, or 4 refer to the J value of contaminating H$_2$ ISM absorption lines. Also, the zero velocity positions of each ISM line are marked by the name of each ion and wavelengths at which geocoronal H I, O I, and O I* emission is seen are marked with the ⊕ symbol. The O VI and C III absorption between 60 and 165 km s$^{-1}$ is highlighted in black in the display of Fig. 4a.

FIG. 5. – Apparent column density, $N_a(v)$, profiles as a function of LSR velocity for the O VI λ1031.926 absorption from FUSE observations of Mrk 421, and for the halo stars BD +38 2182, and HD 93521 are shown in the upper panel. The Mrk 421 O VI profile has been binned to 5 pixels (10 km s$^{-1}$), while the BD +38 2182 and HD 93521 spectra have been binned to two pixels (4 km s$^{-1}$). Note the large breadth of the O VI profile for the extragalactic line of sight compared to the halo star lines of sight. The lower panel shows for Mrk 421 $N_a(v)$ for the O VI λ1031.926 absorption (solid line) and for the C III λ977.020 absorption scaled by 10 times (dotted line). The C III profile is binned to 10 pixels (20 km s$^{-1}$). The vertical solid line at 60 km s$^{-1}$ is the velocity where the C III absorption transitions from very strong to very weak. The weak wings of O VI and C III absorption for $v_{LSR}$ = 60 to 165 km s$^{-1}$ reveals that $N_a$(O VI)/$N_a$(C III) in the positive velocity absorption wing is 10±3 between 60 and 120 km s$^{-1}$.

FIG. 6. - Absorption line profiles for the $z = 0.01$ ($cz = 3022$ km s$^{-1}$) IGM system. The observed flux ($10^{-14}$ erg cm$^{-2}$ s$^{-1}$ Å$^{-1}$) is shown as a function of $cz$ in km s$^{-1}$ along with the fitted continuum. The species plotted include: H I Ly $\alpha$ , H I Ly $\beta$ (blended with ISM C II $\lambda 1036.337$), H I Ly $\gamma$ (not detected), C III $\lambda 977.020$ (not detected but strongly blended with ISM H$_2$ R(1) $\lambda 986.796$.) , C II $\lambda 1036.337$ (not detected) , N V $\lambda 1238.821$ (not detected, O VI $\lambda 1031.926$, (not detected) O VI $\lambda 1037.617$ (strongly blended with ISM Ar I 1048.220). All H$_2$ lines have been fitted and are shown as part of the continuum. The rotational levels from which the H$_2$ absorption lines occur are marked above the zero flux level for each panel at the appropriate velocity of the absorption. Blends from various other ISM absorptions are also marked in the panels. The O VI $\lambda 1031.926$ measurements are from night-only data to reduce contamination from telluric O I emission. The O VI $\lambda 1031.926$ measurements illustrated are from the LiF1A and from LiF2B segments. A strong detector blemish (highlighted in black) contaminates the O VI $\lambda 1031.926$ LiF1A measurements from 3210 to 3430 km s$^{-1}$.

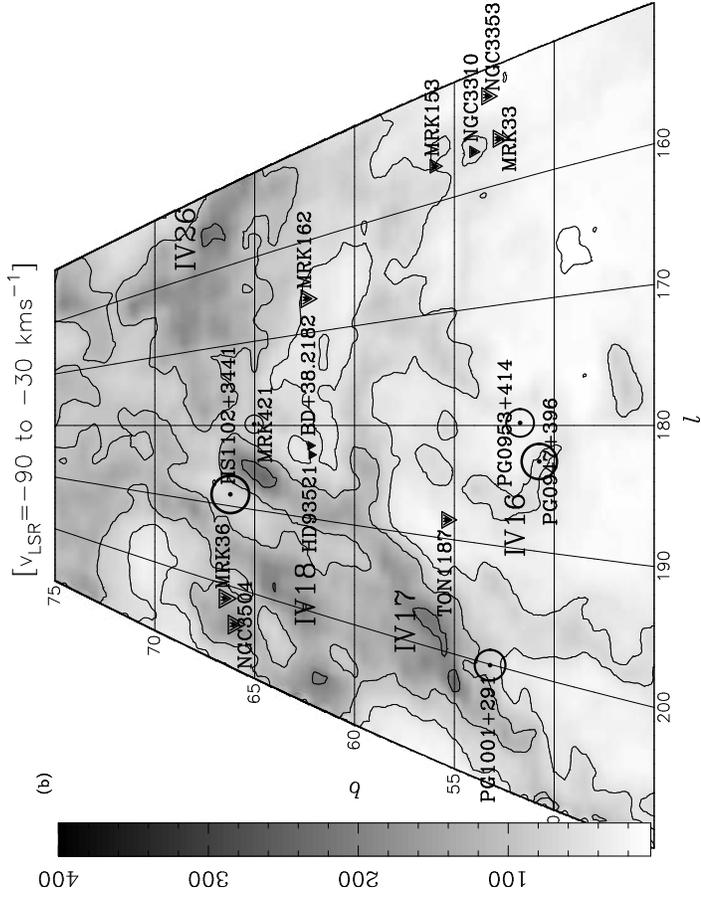

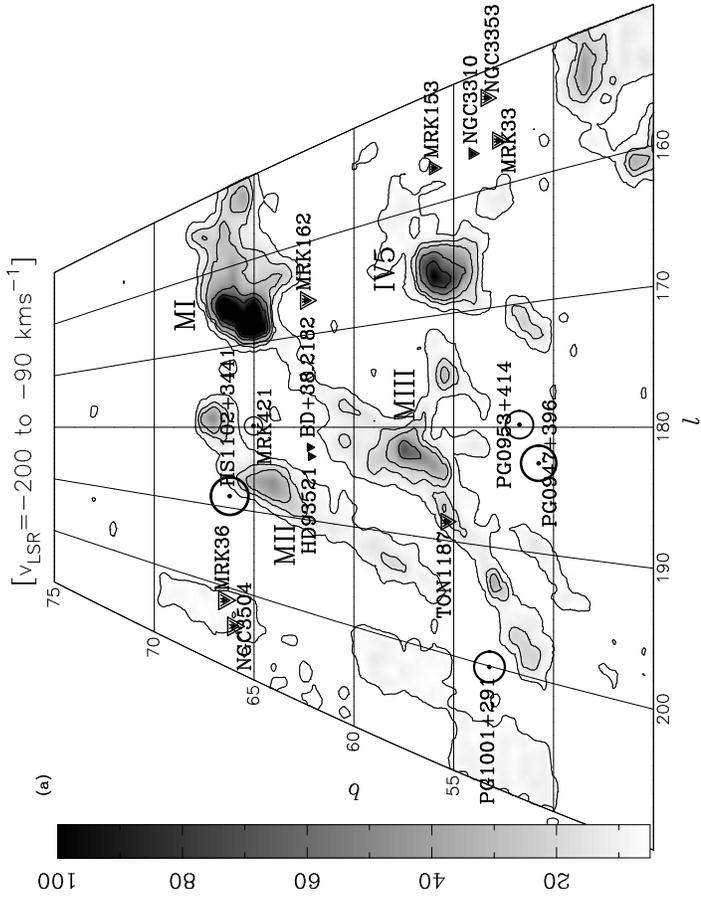

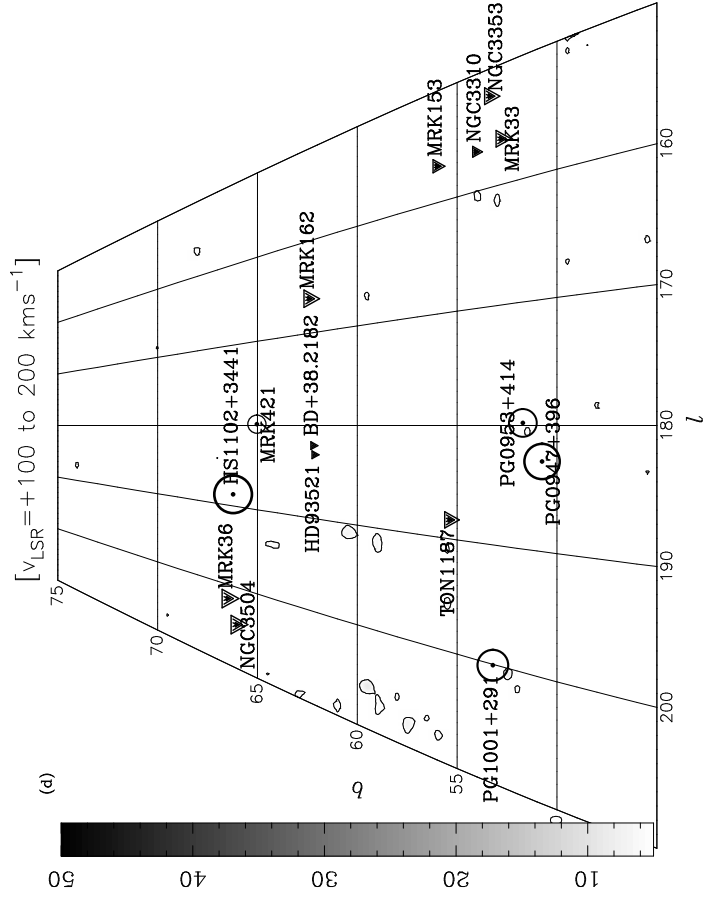

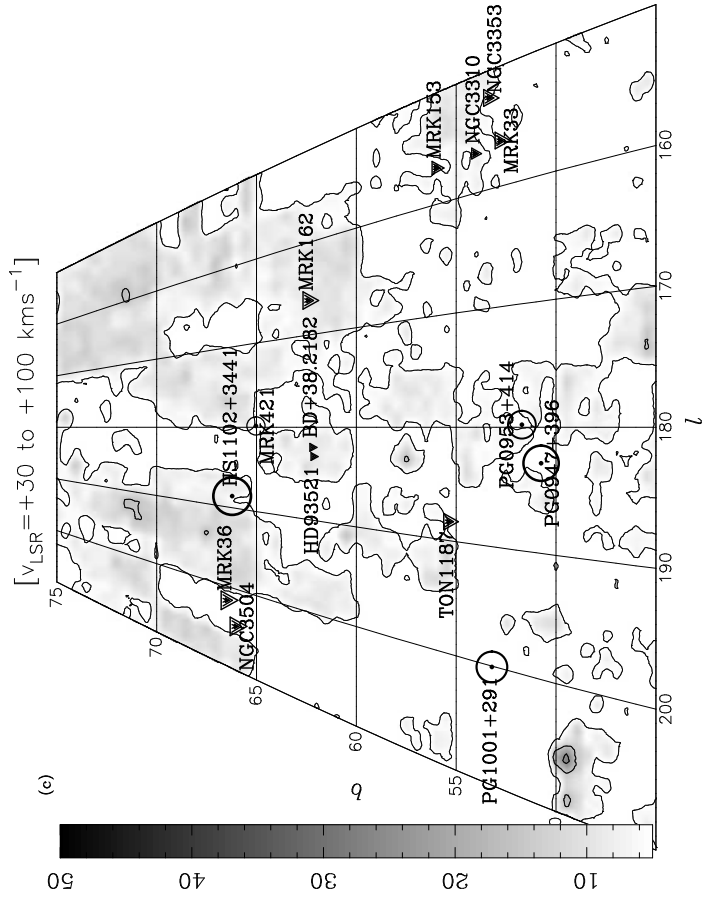

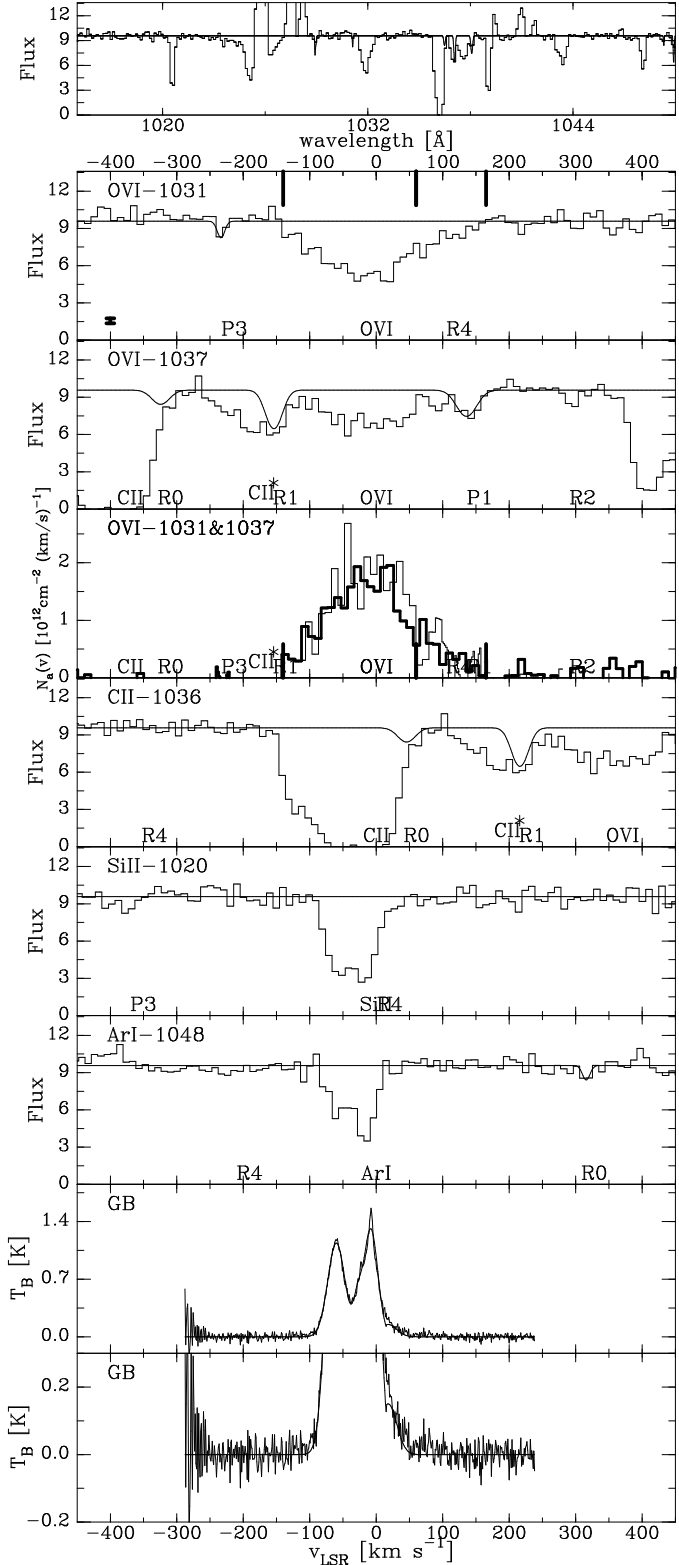

MRK421 (l=179.83 b=65.03)

| nr | vel | amp | FWHM | N(HI) | |
|---|---|---|---|---|---|
| 1 | -60 | 1.14 | 30.2 | 66.7±0.8 | IV26 |
| 2 | -28 | 0.37 | 14.8 | 10.5±1.0 | |
| 3 | -8 | 1.31 | 25.1 | 63.9±1.3 | |
| 4 | 18 | 0.15 | 30.0 | 8.7±0.0 | |

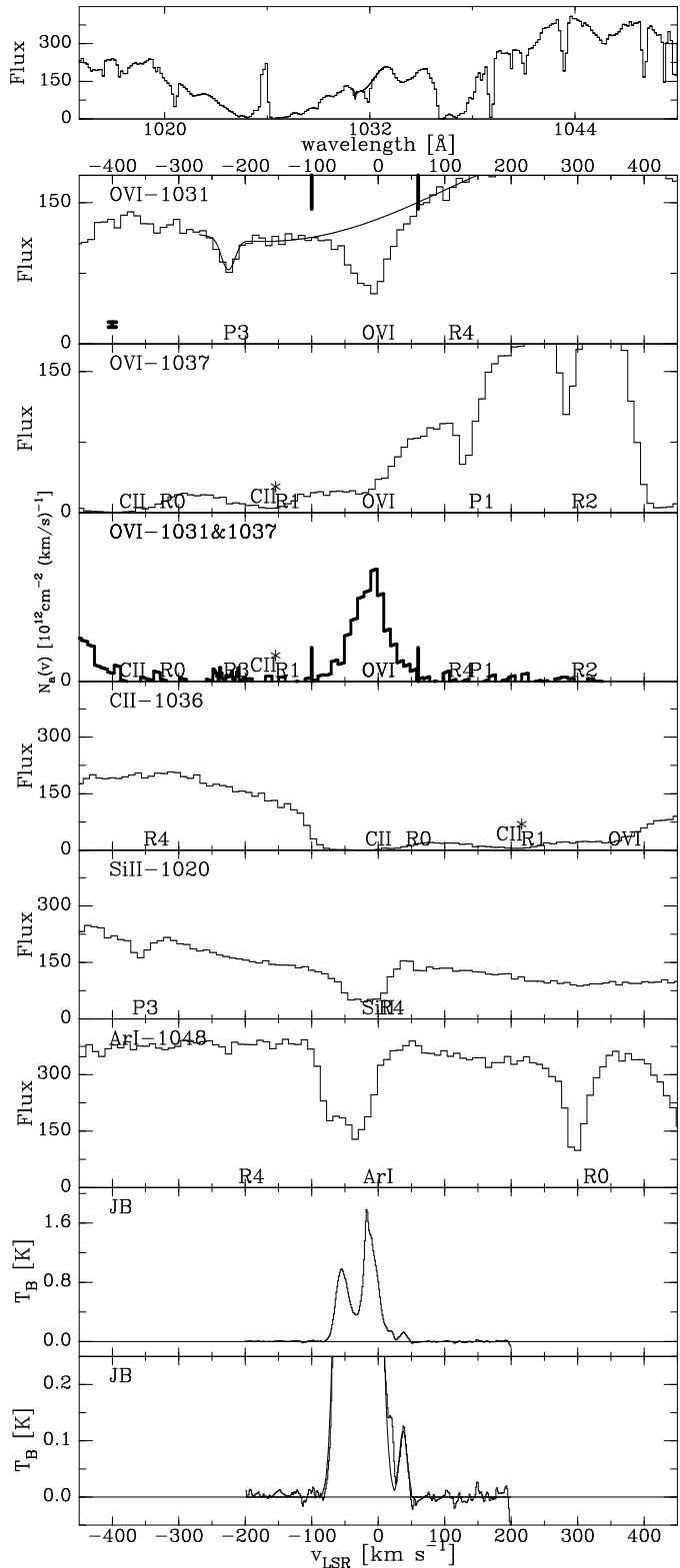

# BD+38.2182 (l=182.16 b=62.21)

| nr | vel | amp | FWHM | N(HI) | |
|---|---|---|---|---|---|
| 1 | -54 | 0.97 | 22.8 | 43.1±0.2 | IV-Arch |
| 2 | -17 | 0.61 | 6.1 | 7.2±0.1 | |
| 3 | -11 | 1.32 | 27.1 | 69.2±0.3 | |
| 4 | 37 | 0.12 | 13.9 | 3.1±0.2 | |

MII [JB: -93 <0.5]

# HD093521 (l=183.14 b=62.15)

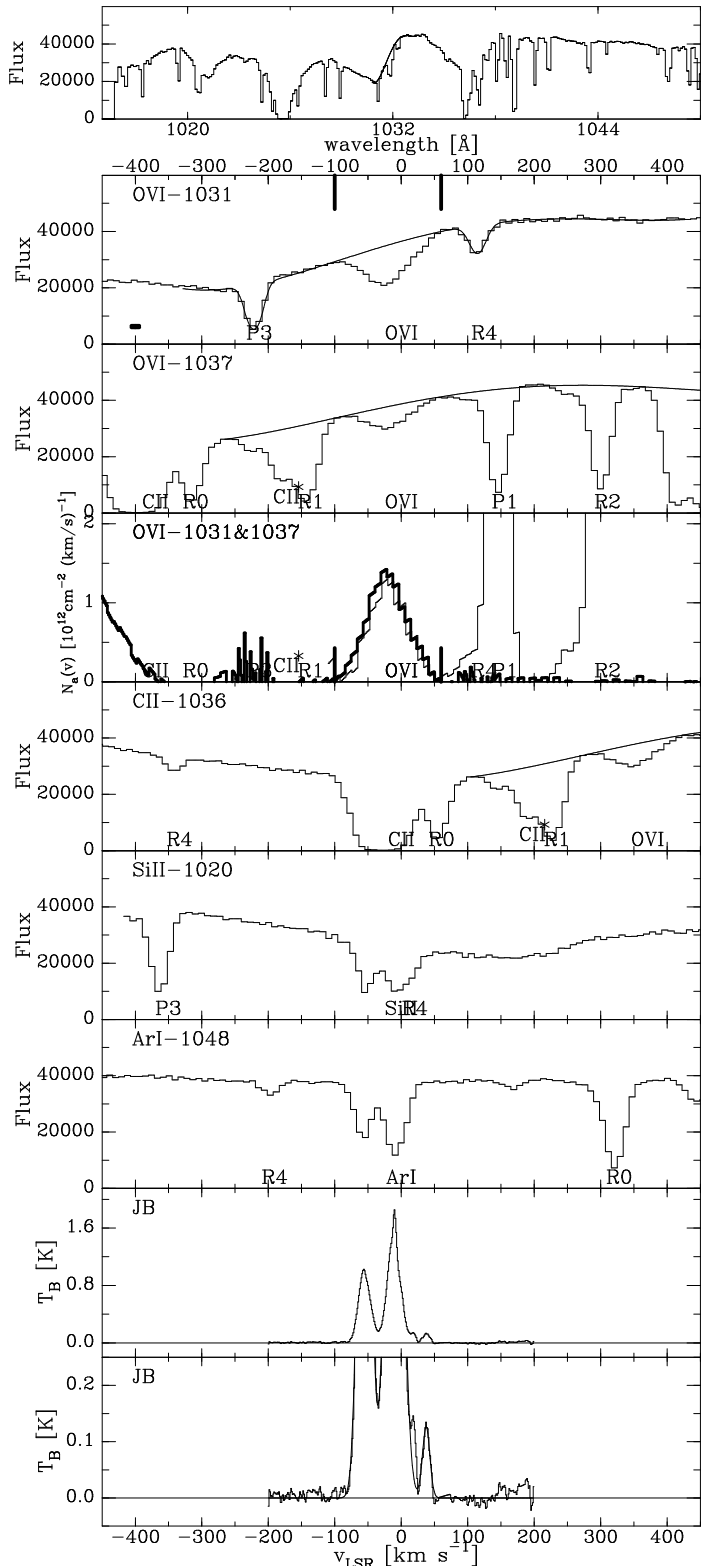

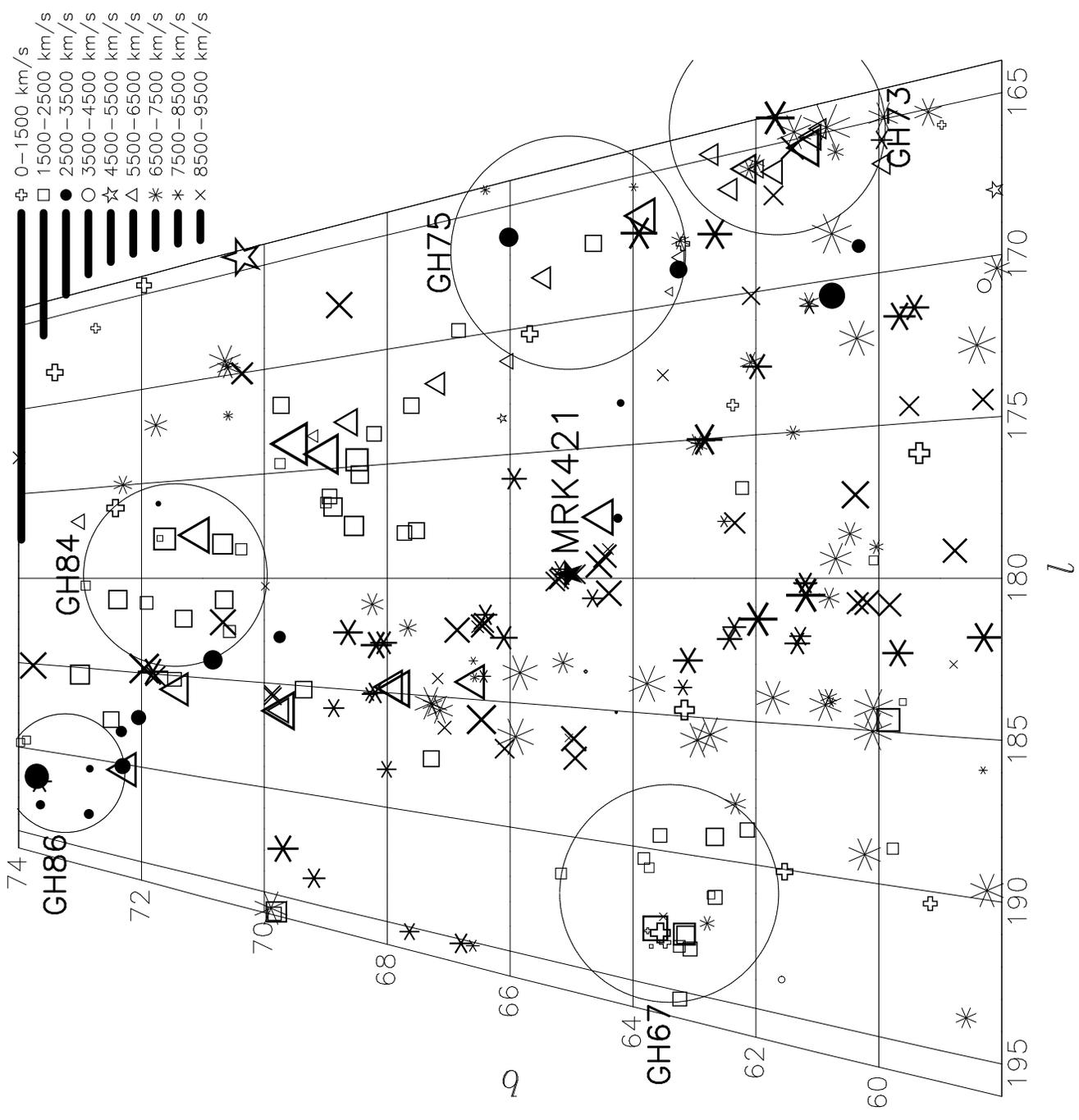

MRK421 (l=179.83 b=65.03)

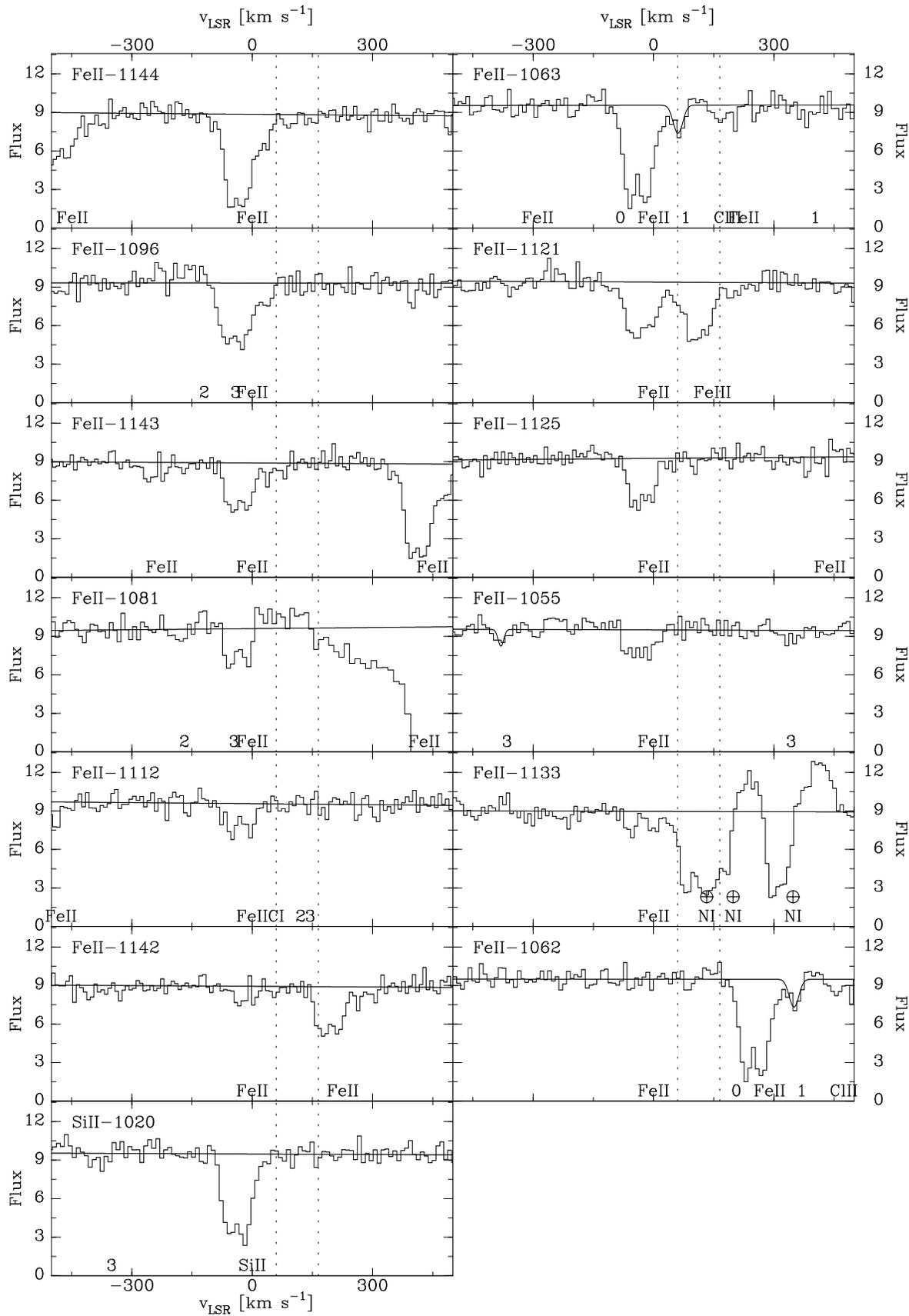

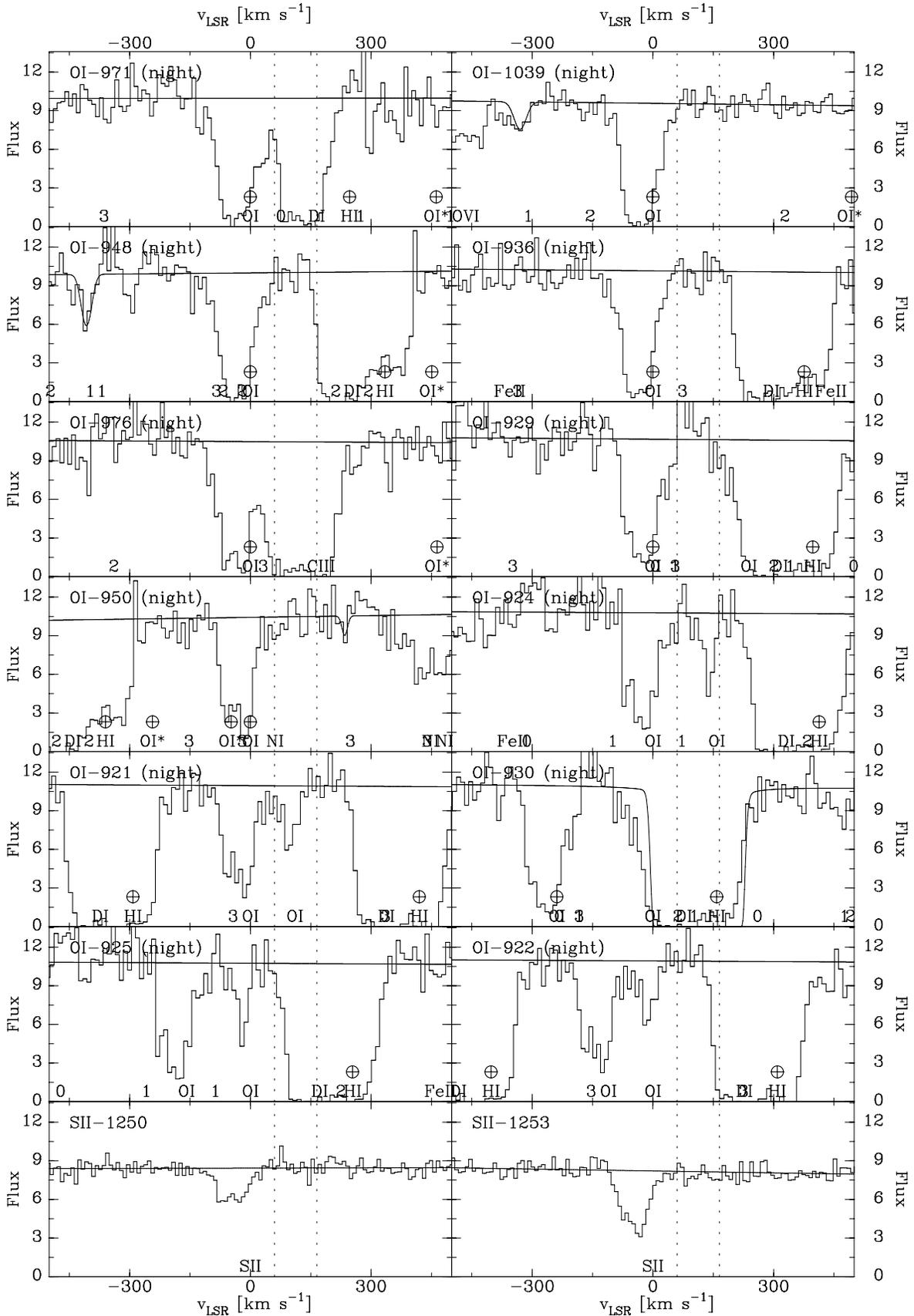

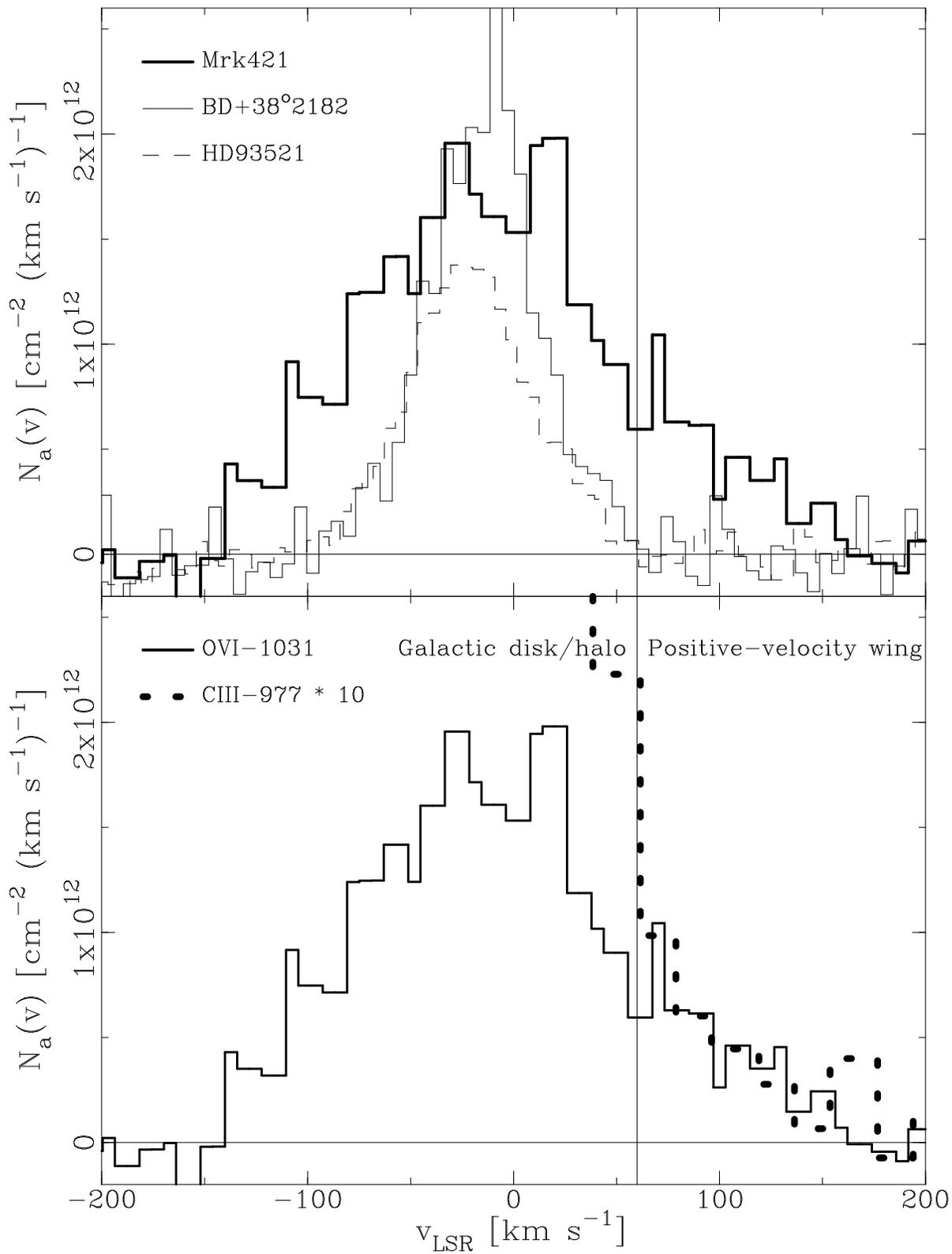

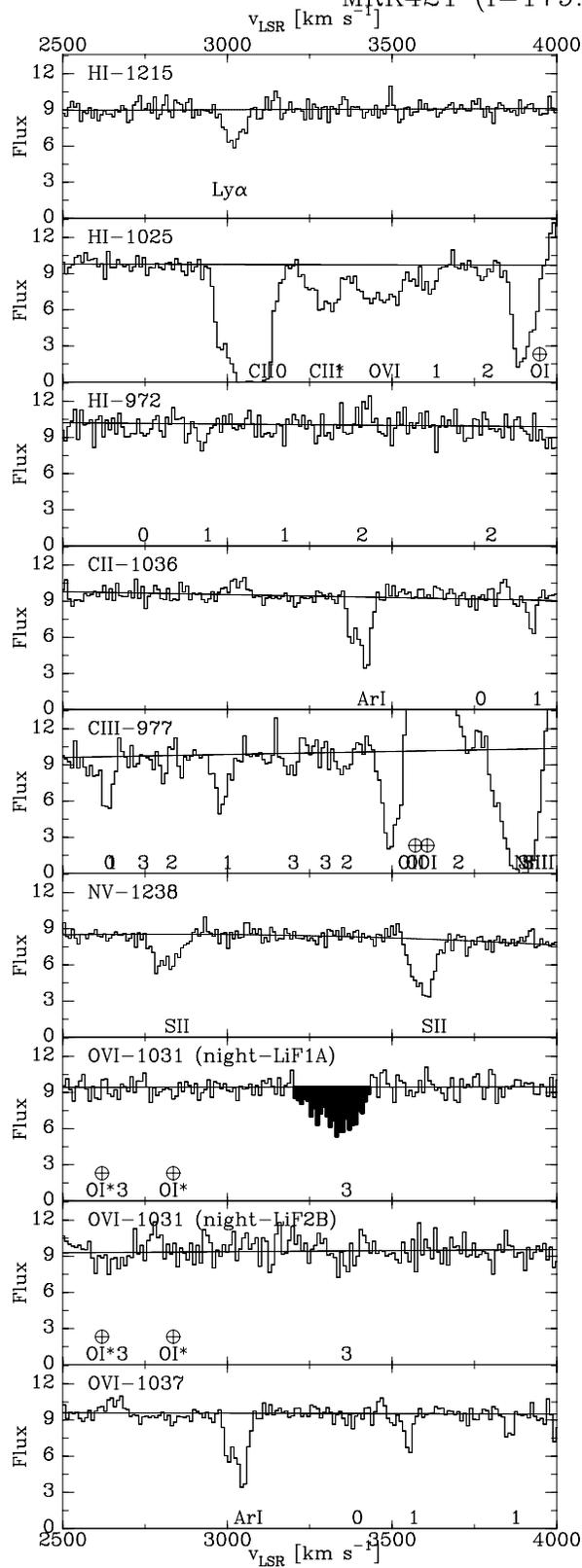

MRK421 (l=179.83 b=65.03)